\documentclass[twocolumn,prd,superscriptaddress,preprintnumbers,nofootinbib]{revtex4}[11pt]
\pdfoutput=1
\usepackage{amsmath,amssymb,graphicx} 
\graphicspath{{figs/}}
\usepackage{epsf,verbatim}
\usepackage{hyperref}
\usepackage{comment}
\usepackage{color}
\usepackage{slashed}
\usepackage{subfigure}
\usepackage[usenames,dvipsnames]{xcolor}
\usepackage{comment}
\usepackage{mdwlist, paralist}
\usepackage{rotating}
\usepackage{multirow}

\def\to{\rightarrow}

\newcommand{\simgt}{\mathrel{\lower2.5pt\vbox{\lineskip=0pt\baselineskip=0pt
           \hbox{$>$}\hbox{$\sim$}}}}
\newcommand{\simlt}{\mathrel{\lower2.5pt\vbox{\lineskip=0pt\baselineskip=0pt
           \hbox{$<$}\hbox{$\sim$}}}}

\newcommand{\squishlist}{
 \begin{list}{$\bullet$}
  { \setlength{\itemsep}{0pt}
     \setlength{\parsep}{3pt}
     \setlength{\topsep}{3pt}
     \setlength{\partopsep}{0pt}
     \setlength{\leftmargin}{1.5em}
     \setlength{\labelwidth}{1em}
     \setlength{\labelsep}{0.5em} } }

\newcommand{\squishlisttwo}{
 \begin{list}{$\bullet$}
  { \setlength{\itemsep}{0pt}
     \setlength{\parsep}{0pt}
    \setlength{\topsep}{0pt}
    \setlength{\partopsep}{0pt}
    \setlength{\leftmargin}{2em}
    \setlength{\labelwidth}{1.5em}
    \setlength{\labelsep}{0.5em} } }

\newcommand{\squishend}{
  \end{list}  }

\newcommand{\beq}{\begin{equation}}
\newcommand{\eeq}{\end{equation}}
\newcommand{\bea}{\begin{eqnarray}}
\newcommand{\eea}{\end{eqnarray}}
\newcommand{\bit}{\begin{itemize}}
\newcommand{\eit}{\end{itemize}}
\newcommand{\OO}{\mathcal{O}}

\newcommand{\f}{\frac}

\newcommand{\secref}[1]{\S\ref{sec:#1}}

\newcommand{\figref}[1]{Fig.~\ref{fig:#1}}

\newcommand{\tabref}[1]{Table~\ref{tab:#1}}
\newcommand{\vev}[1]{ \left\langle {#1} \right\rangle }

\newcommand{\fb}{\,{\rm fb}^{-1}}
\newcommand{\pb}{\,{\rm pb}}

\newcommand{\gev}{{\ \rm GeV}}

\renewcommand{\epsilon}{\varepsilon}
\newcommand{\Br}{{\mathrm{Br}}}
\newcommand{\ie}{{i.e.}}
\newcommand{\eg}{{\it e.g.}}

\begin{document}

\title{
Uncovering light scalars with exotic Higgs decays to $b\bar{b}\mu^+\mu^-$
}
\preprint{YITP-SB-14-53}

\author{David Curtin}
\thanks{dcurtin1@umd.edu}
\affiliation{Maryland Center for Fundamental Physics, University of Maryland, College Park, MD 20742}

\author{Rouven Essig}
\thanks{rouven.essig@stonybrook.edu}
\affiliation{C. N. Yang Institute for Theoretical Physics, Stony Brook University, Stony Brook, NY 11794}

\author{Yi-Ming Zhong}
\thanks{yiming.zhong@stonybrook.edu}
\affiliation{C. N. Yang Institute for Theoretical Physics, Stony Brook University, Stony Brook, NY 11794}

\begin{abstract}
The search for exotic Higgs decays are an essential probe of new physics.  
In particular, the small width of the Higgs boson makes its decay uniquely sensitive to the existence of light hidden sectors.  
Here we assess the potential of an exotic Higgs decay search for $h \to 2X \to b\bar{b}\mu^+\mu^-$ 
to constrain theories with light CP-even $(X = s)$ and CP-odd $(X = a)$ singlet scalars in the mass range of 15 to 60 GeV.
This decay channel arises naturally in many scenarios, such as the Standard Model augmented with a singlet, the two-Higgs-doublet model with a singlet (2HDM+S) -- which includes the Next-to-Minimal Supersymmetric Standard Model (NMSSM) -- and in hidden valley models.
The $2b2\mu$ channel may represent the best discovery avenue for many models. It has competitive reach, and is less reliant on low-$p_T$ $b$- and $\tau$-reconstruction compared to other channels like $4b$, $4\tau$, and $2\tau2\mu$.
We analyze the sensitivity of a $2b2\mu$ search for the 8~and 14~TeV LHC, including the HL-LHC. 
We consider three types of analyses, employing conventional resolved $b$-jets with a clustering radius of $R\sim 0.4$, 
thin $b$-jets with $R=0.2$, and jet substructure techniques, respectively.  
The latter two analyses improve the reach for $m_X\sim 15$~GeV, for which the two $b$-jets are boosted and often merged.  
We find that $\Br(h\to2X\to2b2\mu)$ can be constrained at the few $\times$ $10^{-5}$ level across the entire considered 
mass range of $X$ at the HL-LHC. 
This corresponds to a $1 - 10\%$ reach in $\Br(h\to2X)$ in 2HDM+S models, including the NMSSM, depending on the type of Higgs Yukawa couplings.
\end{abstract}

\maketitle

 \setcounter{equation}{0} \setcounter{footnote}{0}

%%%%%%%%%%%%%%%%%%%%%%%%%%%%%%%%%%%%%%%%%%%%%%%%%%%%%%%%%%%%%%%%%%%%%%
%%%%%%%%%%%%%%%%%%%%%%%%%%%%%%%%%%%%%%%%%%%%%%%%%%%%%%%%%%%%%%%%%%%%%%% 
\section{Introduction}

The discovery of the 125 GeV Higgs boson at the Large Hadron Collider (LHC)~\cite{Aad:2012tfa,Chatrchyan:2012ufa} 
opens up several new experimental frontiers.  The complete characterization of this new particle, including the precise 
measurements of its couplings, searches for Higgs ``siblings'', and searches for non-standard (exotic) decay 
modes~\cite{Curtin:2013fra,Chang:2008cw,Chang:2005ht}, has the great potential to reveal signs of physics beyond 
the Standard Model (SM).  
Among the most exciting possibilities is that the Higgs boson can provide a unique window onto light hidden sectors, consisting of 
particles neutral under the SM gauge groups.  

The Higgs boson is one of only a few SM particles that can couple to new states with an interaction that is (super-)renormalizable.  
In addition, the small decay width of the SM Higgs, dominated by the bottom Yukawa coupling, means that a small, $\mathcal{O}(0.01)$,  renormalizable coupling of the Higgs to a new, light state can lead to an exotic Higgs decay branching fraction of $\mathcal{O}(1)$.  
This makes exotic Higgs decays a prime experimental target.  
In many cases, these exotic decays need to be searched for explicitly as they may otherwise escape detection.  
In particular, measurements of the Higgs couplings to SM states only constrains the Higgs branching ratio to non-SM 
states to $\lesssim 60\%$~\cite{CMS-PAS-HIG-12-045,ATLAS-CONF-2013-034}.  
Thus a large branching ratio to beyond SM particles is still viable. 
For a detailed survey of promising exotic decay modes and their theoretical motivations we refer the reader to~\cite{Curtin:2013fra}. 

One interesting category of exotic Higgs decays contains final states with four SM fermions and no missing energy: 
$h \to XX' \to 2 f 2 f'$, where $X$ and $X'$ are on-shell, and we here assume that they are the same particle, $X=X'$.\footnote{We use 
the shorthand, for example,  `$2f$' or `$4f$' to denote $f\bar{f}$ of $f\bar{f}f\bar{f}$, respectively.}
Generically, the couplings of $X$ determine the optimal search strategy.  If $X$ is a dark photon, i.e.~the mediator of a new, broken 
$U(1)$ gauge theory which kinetically mixes with the SM hypercharge 
gauge boson~\cite{Holdom:1985ag,Galison:1983pa,Dienes:1996zr}, then the couplings of $X$ to SM particles are gauge-ordered, 
i.e.~the $X$ couplings are related to the SM $Z$-boson and photon couplings to SM fermions.  
In this case, the $X$ has an $\mathcal{O}(1)$ branching fraction to light leptons, making $h \to 4\ell$ the best discovery 
channel~\cite{Gopalakrishna:2008dv,Jaeckel:2012yz,Davoudiasl:2012ag,Davoudiasl:2013aya,Chang:2013lfa,Curtin:2013fra,Falkowski:2014ffa,Cline:2014dwa,Hoenig:2014dsa,Curtin:2014cca}. 
On the other hand, if $X$ is a CP-odd\footnote{In this study, we will only consider CP-conserving Higgs sectors.} scalar ($a$) or a CP-even scalar ($s$), it generically inherits its couplings from the SM Higgs sector.  This means that the couplings of $X$ to the SM fermions are typically Yukawa-ordered, so that its largest branching fraction is to the heaviest fermion that is kinematically accessible.  For this reason, previous LHC studies have extensively focused on the decay channels $h\to 4b$~\cite{Ellwanger:2003jt, Ellwanger:2005uu, Cao:2013gba, Cheung:2007sva, Carena:2007jk, Kaplan:2011vf} and $h\to 2 b 2\tau$~\cite{Adam:2008aa, Carena:2007jk} for $m_X>2m_b$, 
$h\to 4\tau$~\cite{Belyaev:2008gj, Englert:2011iz} and $h\to 2\tau2\mu$~\cite{Lisanti:2009uy,Abazov:2009yi} 
for $2m_\tau< m_X < 2m_b$, and $h\to 4\mu$~\cite{Abazov:2009yi,Chatrchyan:2012cg,Chatrchyan:2011hr,CMS:2013lea} 
for $2m_\mu< m_X < 2m_\tau$.  
These searches are motivated in the context of, for example, the SM with a singlet (see e.g.~\cite{Curtin:2013fra}); the two-Higgs-doublet model with an additional singlet (2HDM+S, see e.g.~\cite{Chang:2005ht,Curtin:2013fra}), including the next-to-minimal supersymmetric standard model (NMSSM)~\cite{Dermisek:2005ar,Dermisek:2007yt,Ellwanger:2009dp}; the minimal supersymmetric standard model (MSSM) with a singlet~\cite{Chang:2005ht}; as well as many hidden valley models~\cite{Strassler:2006qa, Strassler:2006im, Strassler:2006ri,Strassler:2008bv}. 

In this paper we propose a new search channel, $h\to2b2\mu$, as a promising discovery avenue for Higgs decays to light scalars with a mass above $2m_b$.  As we will see below, this channel represents a compromise between the dominant but difficult $4b$ and $2b2\tau$ channels, and the spectacular but very rare $4\mu$ channel.  In~\cite{Curtin:2013fra}, two scenarios for realizing this decay via intermediate on-shell states were considered: $h \to Z a$ (see also~\cite{Coleppa:2014hxa}) and $h \to XX$ with $X=s$ or $X=a$. Sensitivity to the latter 
scenario was only estimated at parton-level.  
Here we expand on this estimate and provide a more detailed and comprehensive collider study for $h\to 2X \to 2b2\mu$ at the LHC. We also discuss how the projected sensitivity compares to the results of previous collider studies in the $4b$, $4\tau$, $2\tau2\mu$,  and $2b2\tau$ channels.

The paper is organized as follows. We first review the theoretical motivation for a search of $h\to 2X \to 2b2\mu$ in \S\ref{sec:TheoreticalMotivation}. We then discuss the sensitivity projections of this channel at the LHC~8 and LHC~14 in \S\ref{sec:ConstraingExclusionyPotential}, discuss and compare these with existing sensitivity projections for other decay 
modes in \S\ref{sec:discussion}, and finally conclude in \S\ref{sec:conclusion}. 
Some details about fake-lepton background estimates are included in an Appendix.
 
%%%%%%%%%%%%%%%%%%%%%%%%%%%%%%%%%%%%%%%%%%%%%%%%%%%%%%%%%%%%%%%%%%%%%%
%%%%%%%%%%%%%%%%%%%%%%%%%%%%%%%%%%%%%%%%%%%%%%%%%%%%%%%%%%%%%%%%%%%%%
\section{Theoretical Motivation}
\label{sec:TheoreticalMotivation}

In this section, we discuss a non-exhaustive set of models that contain the $h\to 2b2\mu$ decay.  
We only consider the SM with a singlet and the 2HDM+S models, as well as the NMSSM in particular. 
In these models, the $h$ decays to an intermediate on-shell scalar, which is either CP-even (and denoted by $s$) or 
CP-odd (and denoted by $a$), i.e.~we consider $h\to ss$, or $h\to aa$.  
We will not consider other models that can lead to this decay.  
It is also possible that the Higgs decays to two scalars with {\it different} masses and/or couplings, e.g.~$h\to ss'$ or $h\to a a'$, 
where $s$ and $a$ ($s'$ and $a'$) have large branching ratios to $b\bar b$ ($\mu^+ \mu^-$). 
We do not consider this possibility in detail here. However, if it was realized, the $2b2\mu$ channel would obviously offer the best sensitivity to the total exotic Higgs decay branching fraction.

\subsection{Standard Model plus a Singlet (SM+S)}\label{subsec:SMS}
A minimal modification of the SM is to add one real scalar singlet $S$ (``SM+S") that mixes with the SM Higgs after electroweak symmetry breaking (EWSB). 
We take the renormalizable potential for the SM+S to be  
\beq
V(H, S)=-\mu^2 |H|^2+\lambda |H|^4 -\frac{1}{2} \mu_S^2 S^2 + \f{1}{4}\lambda_S S^4 +\f{1}{2} \kappa S^2|H|^2,
\label{eq:SMSV}
\eeq
where $H$ is the SM Higgs doublet.  We choose the couplings in such a way that $S$ gets a nonzero vacuum expectation 
value (vev), 
breaking the $\mathbb{Z}_2$ symmetry $S\to -S$ and allowing $H$ and $S$ to mix after EWSB.   
The surviving real degrees of freedom after EWSB consist of two neutral CP-even scalars, $h$ and $s$.  We take $h$ to be 
the 125 GeV Higgs boson and $s$ to satisfy $m_s < m_h/2$.  The branching ratio Br($h\to ss$) can easily be 
sizable~\cite{Curtin:2013fra}, and the mixing between $h$ and $s$ allows $s$ to decay to SM particles with branching 
ratios inherited from the $h$ decay to SM particles. 
This means that the decays to the heaviest SM fermions with a mass less than $m_s/2$ dominate: in the case of $m_s > 2 m_b$, the dominant decay is $b \bar b$.

We introduce
\beq\label{eq:epsilonmub}
\epsilon_{\mu b}\equiv \frac{\Br (s \to \mu^+ \mu^-)}{\Br(s \to b \bar b)} \approx \frac{m_{\mu}^2}{3 m_{b}^2}\approx 2\times 10^{-4}
\eeq
to characterize the couplings of $s$ to muons. The small value of $\epsilon_{\mu b}$ explains the hierarchical structure 
of the $s$ branching ratios to $4\mu$, $2b 2\mu$, and $4b$.  At leading order, and ignoring phase space corrections, the Higgs 
branching ratios satisfy 
\bea\label{eq:BR-SMS}
\Br(h\to 2s \to 4\mu)&=&\frac{\epsilon_{\mu b}}{2} \Br (h\to 2s\to 2b2\mu)\nonumber\\
&=&\epsilon_{\mu b}^2 \Br(h\to 2s \to4b)\,.
\eea
Precise values, including QCD corrections that are calculated following~\cite{Djouadi:2005gj, Djouadi:2005gi}, are shown in   \tabref{smsbr}. 

Assuming that the Higgs is produced with SM rates, and that $\Br(h\to 2s)=10\%$, one can estimate that $\mathcal{O}(20)$ $h\to 2s \to 2b2\mu$  events could be observed from  gluon-gluon fusion (ggF) Higgs production at the LHC Run I (compared to zero $h\to 2s \to 4\mu$ events). 
While this is much less than the few hundred $h\to 2s \to 4b$ events expected from associated  production, the backgrounds for a $W(h\to4b)$ search are very challenging. As we discuss in \secref{discussion}, $2b2\mu$ provides complementary information to the the usual $4b$ channel for an SM+S-like scenario, and may be superior, depending on how well 
relatively soft $b$-jets can be reconstructed.  

\begin{table}[t!]
\begin{center}
\begin{tabular}{|l|c|}
\hline
Final State & $\Br(h\to2s\to2f2f')/ \Br(h\to2s)$
\\ \hline
$b \bar b b \bar b$ &  $0.77$
\\ \hline
$b \bar b \tau^+ \tau^-$ & $0.10$
\\ \hline
$\tau^+ \tau^- \tau^+ \tau^-$ & $3.5 \times 10^{-3}$
\\ \hline
$b \bar b \mu^+ \mu^-$ &   $3.7 \times 10^{-4}$
\\ \hline
$\tau^+ \tau^- \mu^+ \mu^-$& $2.5 \times 10^{-5}$
\\ \hline
$\mu^+ \mu^- \mu^+ \mu^-$ & $4.5 \times 10^{-8}$
\\ \hline
\end{tabular}
\end{center}
\caption{
$\mathrm{Br}(h \to 2s \to 2f 2f')/\Br(h\to 2s)$ in the SM+S model, with $m_s = 40 \gev$. 
These numbers are relatively constant across the mass range $15 \gev \leq m_s \leq 60 \gev$. 
}
\label{tab:smsbr}
\end{table}

\subsection{2HDM+S}\label{2HDMS}

\begin{table*}[t!]
   \centering
\begin{tabular}{ |p{1cm}ccccc| }
\hline
  & Couplings & I & II (NMSSM-like) & III (Lepton specific) & IV (Flipped) \\ \hline
\multirow{5}{*}{$h$} 
& $g_{h VV}$ & $\sin (\beta-\alpha)$ &$ \sin (\beta-\alpha)$ &$ \sin (\beta-\alpha)$ &$ \sin (\beta-\alpha)$ \\
 & $g_{h t \bar t}$ & $ \cos\alpha/\sin\beta$& $ \cos\alpha/\sin\beta$ &$ \cos\alpha/\sin\beta$&$ \cos\alpha/\sin\beta$ \\
 &  $g_{h b \bar b}$ & $ \cos\alpha/\sin\beta$&$- \sin\alpha/\cos\beta$&$ \cos\alpha/\sin\beta$&$- \sin\alpha/\cos\beta$ \\
 &  $g_{h \tau \bar \tau}$ & $ \cos\alpha/\sin\beta$&$- \sin\alpha/\cos\beta$&$- \sin\alpha/\cos\beta$&$ \cos\alpha/\sin\beta$ \\  
  \hline
 \multirow{5}{*}{$H^0$} 
& $g_{H^0 VV}$ & $\cos (\beta-\alpha)$ &$ \cos (\beta-\alpha)$ &$ \cos (\beta-\alpha)$ &$ \cos (\beta-\alpha)$ \\
 & $g_{H^0 t \bar t}$ & $ \sin\alpha/\sin\beta$& $ \sin\alpha/\sin\beta$ &$ \sin\alpha/\sin\beta$&
$ \sin\alpha/\sin\beta$ \\
 &  $g_{H^0 b \bar b}$ & $ \sin\alpha/\sin\beta$&$ \cos\alpha/\cos\beta$&$ \sin\alpha/\sin\beta$&$ 
\cos\alpha/\cos\beta$ \\
 &  $g_{H^0 \tau \bar \tau}$ & $ \sin\alpha/\sin\beta$&$\cos\alpha/\cos\beta$&$ \cos\alpha/\cos\beta$&
$ \sin\alpha/\sin\beta$ \\  
 \hline

\multirow{4}{*}{$A$} 
 & $g_{A VV}$ & 0&0&0&0 \\
 & $g_{A t \bar t}$  & $ \cot \beta$ &  $ \cot \beta$ &$ \cot \beta$ &$ \cot \beta$ \\
 & $g_{A b \bar b}$ & $-  \cot \beta$ & $ \tan \beta$ &$- \cot \beta$ & $ \tan \beta$ \\
 & $g_{A \tau \bar \tau}$ & $-  \cot \beta$ & $ \tan \beta$& $ \tan \beta$ &$- \cot \beta$\\ \hline
\end{tabular}
   \caption{Couplings of the mass eigenstates of the neutral CP-even scalars $h$ and $H^0$, and CP-odd scalar $A$ in the four types of 2HDM with a $\mathbb{Z}_2$ symmetry. The table follows the convention of \cite{Craig:2012pu}. All couplings are normalized to those of the SM Higgs, and only the coupling to the heaviest SM fermion with a particular set of quantum numbers is shown. 
   Here $\tan \beta \equiv  \vev{H_2}/\vev{H_1}$ and the mixing angle $\alpha \in (-\pi/2, \pi/2)$ defines the admixture of $H_{1,2}$ that make up the mass eigenstates $h, H^0$. 
   In the 2HDM+S setup, the couplings of the singlet-like pseudoscalar $a$ are identical to the couplings of $A$, up to an overall mixing angle. The couplings of the singlet-like scalar $s$ can be obtained (again up to an overall mixing angle) from the $h$-couplings by replacing $\alpha \to \alpha'$, where the free parameter $\alpha'$ defines the mixture of $H_{1,2}$ that mixes with $s$ (see~\cite{Curtin:2013fra} for details).  The couplings listed here can be used for the calculation of the singlet branching ratios in the 2HDM+S, as additional mixing angles drop out.} 
   \label{tab:2HDMcoupling}
\end{table*}

The simple SM+S set-up can be generalized to a two-Higgs-doublet model (``2HDM'') (see \cite{Craig:2012pu, Craig:2013hca, Chen:2013kt} for recent reviews) with an additional complex singlet (``2HDM+S'').  We refer the reader to~\cite{Curtin:2013fra} for 
a recent detailed review of the Higgs phenomenology in the 2HDM+S model, and only discuss the most salient features here. 
Much of the parameter space of these models remains unexplored by existing experimental data.
(Note that the unaugmented 2HDM can also generate exotic higgs decays of the $h\to aa$ type, see e.g. \cite{Bernon:2014nxa}.)

We start by considering the four 2HDM models in which SM fermions with the same quantum numbers couple to only one Higgs field (this 
avoids large flavor-changing neutral currents).  After EWSB, the  neutral physical states of the 2HDM sector consist of two CP-even scalars $(h, H^0)$
and one CP-odd scalar $(A)$.  Their couplings to the SM fermions and gauge bosons are summarized in \tabref{2HDMcoupling}.  
The couplings between the CP-odd scalar and SM fermions are controlled by the value of  $\tan \beta \equiv \vev{H_2}/\vev{H_1}$, 
where $H_1$ and $H_2$ are the two Higgs doublets, as well as the Yukawa coupling type. The couplings of the two CP-even scalars to fermions additionally depend on the mixing angle $\alpha$, which dictates the admixture of $H_1$ and $H_2$ that make up the mass eigenstates $h, H^0$. In the decoupling limit, $m_{A} \to \infty$, $\alpha \to \beta-\pi/2$. 
Higgs coupling measurements, with $h$ identified as the discovered 125 GeV state, already place significant constraints on $\alpha$ and $\beta$, see e.g.~\cite{Coleppa:2013dya, Craig:2013hca, Chang:2013ona, Barger:2013ofa}. 

\begin{figure*}[t!]
   \centering
   \includegraphics[width=0.48\textwidth]{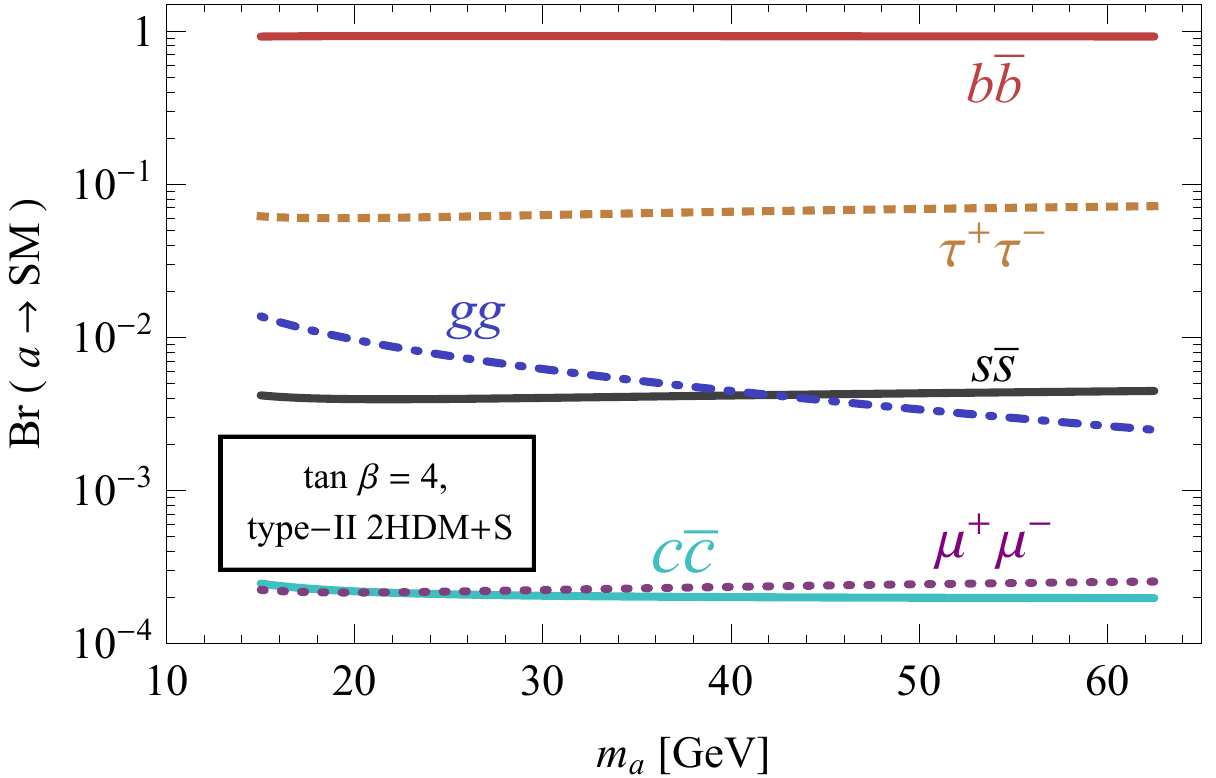}~~
   \includegraphics[width=0.48\textwidth]{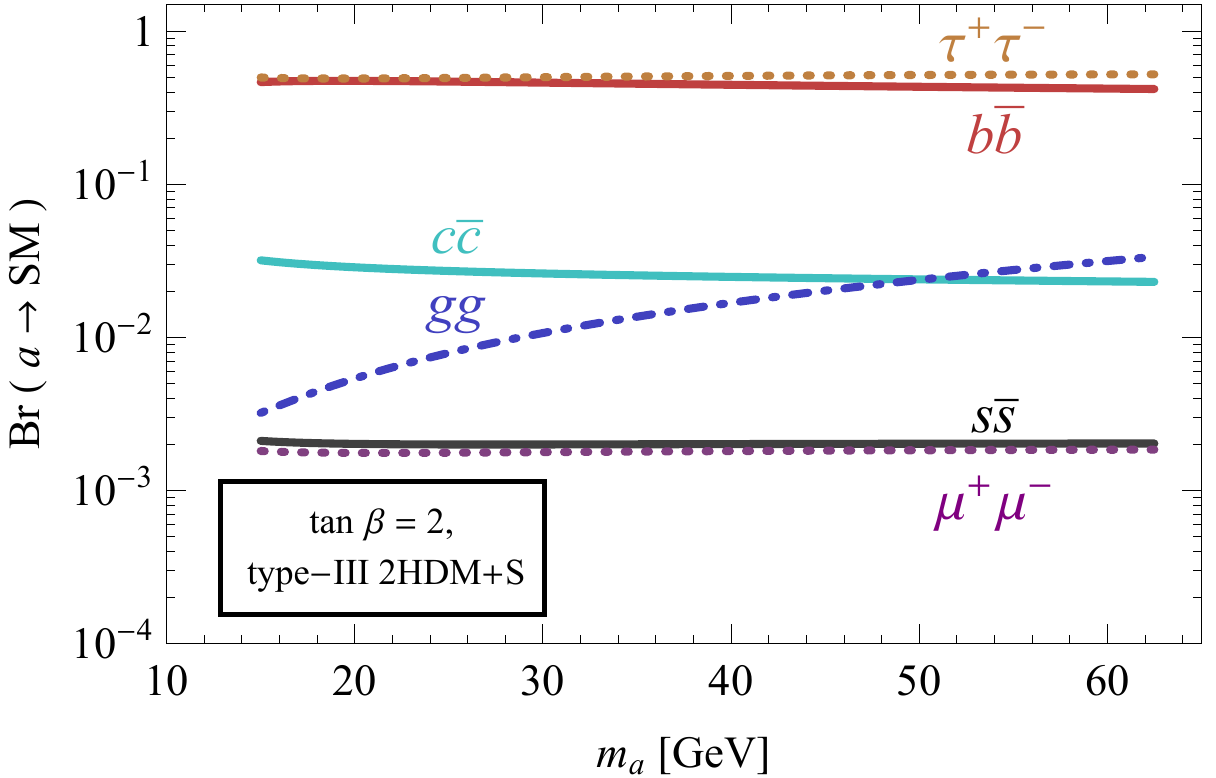}      
   \caption{Branching ratios of a CP-odd scalar $a$ in a 2HDM+S type-II model  with $\tan \beta=4$ ({\it left}) and 
   a type-III model with $\tan \beta=2$ ({\it right}).  
   For the type-II model, $\Br(h \to2a \to 2b2\mu) / \Br(h\to 2a) \simeq 4.0\times 10^{-4}$ for $15 \gev < m_a < m_h/2$, which is very similar to the SM+S scenario, see \tabref{smsbr}. 
For the type-III model, $\Br(h \to 2a\to 2b2\mu)/\Br(h\to 2a)\simeq 1.6\times 10^{-3}$ for $15 \gev < m_a < m_h/2$, which is 
 enhanced by about a factor of $4$ compared to the SM+S in \tabref{smsbr}.}
   \label{fig:Br}
\end{figure*}

We now add to the 2HDM model a complex singlet $S$, which has a small mixing with $H_1$ and/or $H_2$. This leads to two additional physical states that are mostly singlet-like but inherit interactions to the SM fermions from their mixing with the Higgs doublets: one CP-even scalar, $s$, and one CP-odd scalar, $a$. The couplings of $a$ are entirely inherited by mixing with the pseudoscalar state $A$, and can be read off from \tabref{2HDMcoupling}, up to an overall mixing angle rescaling. On the other hand, the couplings of $s$ depend on the admixture of $H_1$ and/or $H_2$ that mixes with $s$. This admixture can be defined, in analogy to the doublet mass eigenstates, by an effective mixing angle $\alpha'$. The couplings of $s$ can then be obtained from the $h$-couplings in \tabref{2HDMcoupling} by replacing $\alpha \to \alpha'$, again up to an overall rescaling. (See \cite{Curtin:2013fra} for more details.) Note that the singlet masses $m_a$ and $m_s$, as well as $\alpha'$, are in principle completely free parameters of the theory, independent of $\tan \beta$ and $\alpha$.  If the overall mixing between the singlet and doublets is small enough, constraints on $\alpha$ and $\beta$ in 2HDM's also apply to 2HDM+S, but $\alpha'$ can take on any value. 
For a more detailed discussion on coupling constraints in the 2HDM+S see~\cite{Chen:2013jvg}.

\begin{figure*}[t!]
   \centering
   \includegraphics[width=0.48\textwidth]{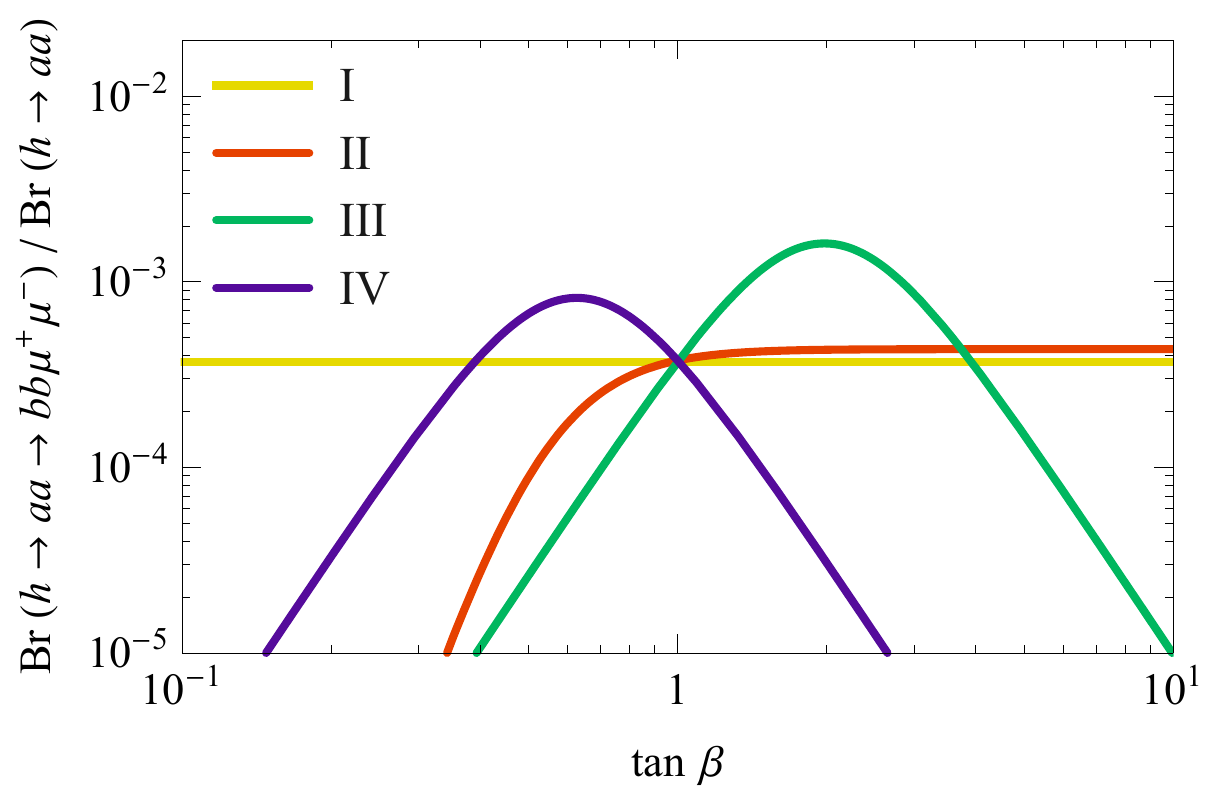} 
   \includegraphics[width=0.48\textwidth]{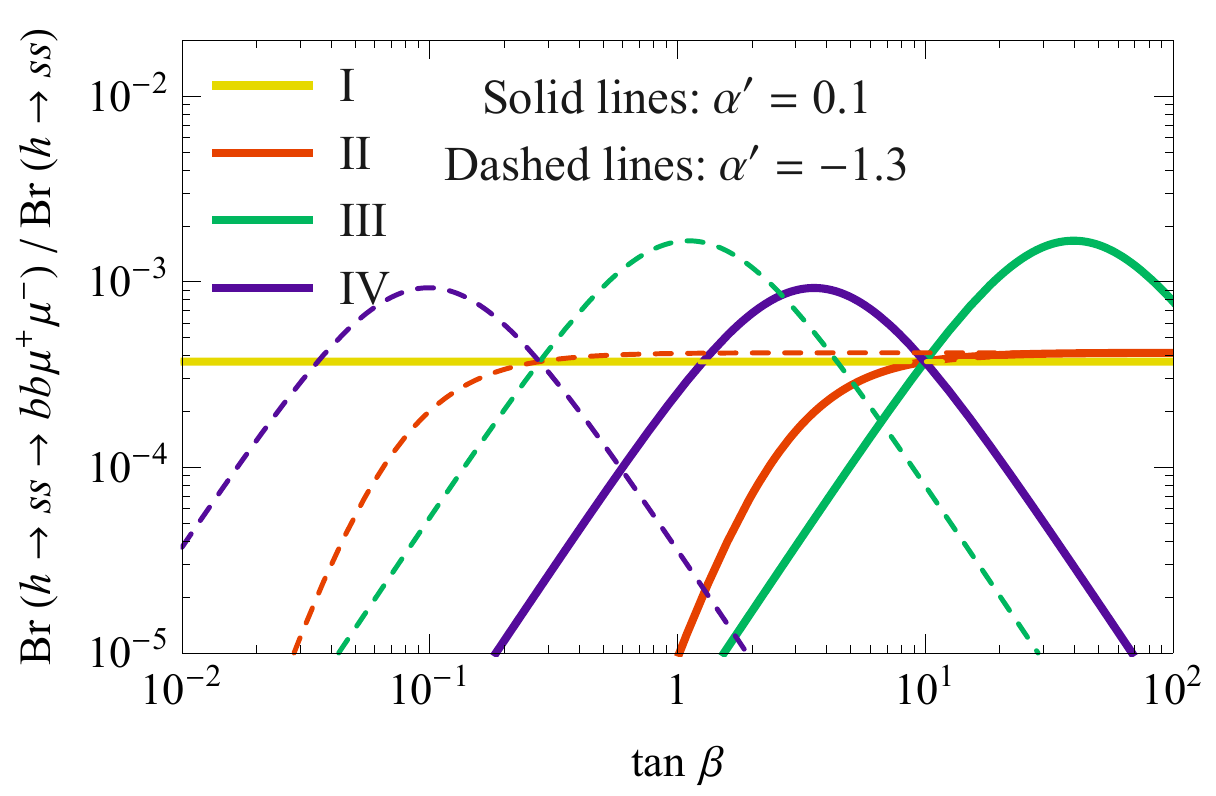} 
      \caption{$\Br(h \to2a \to 2b2\mu) / \Br(h\to 2a)$ of a CP-odd scalar, $a$, ({\it left}) and CP-even scalar, $s$, ({\it right}) 
      in 2HDM+S with a fixed mass $m_{a(s)}=40 \gev$.}
   \label{fig:Br-vs-tanbeta}
\end{figure*}

The general 2HDM+S setup generates a rich phenomenology.  
In particular, the simple scaling of the branching ratios given in Eqs.~(\ref{eq:epsilonmub}) and (\ref{eq:BR-SMS}) does not hold in all regions of parameter space.  Of interest to us here are scenarios for which the Higgs decay branching ratio to $2b2\mu$ is enhanced  compared to its value in the SM+S model.  
\figref{Br} shows the branching ratios of the CP-odd scalar $a$  as a function of  $m_a$ for a particular choice of $\tan \beta$ in the type-II ({\it left}) and type-III ({\it right}) 2HDM. 
While the type-II case shown provides an example with a very similar phenomenology to the SM+S model in \S\ref{subsec:SMS}, 
the type-III case shown features significantly larger Br($h\to aa \to 2b2\mu$).

Above the $b\bar b$ threshold, the relevant branching ratios depend only weakly on mass. It is therefore instructive to look at  Br($h\to2a(s)\to2b2\mu$) as a function of $\tan \beta$ (and $\alpha'$) for the four 2HDM model types.
The {\it left} plot in \figref{Br-vs-tanbeta} shows $\Br(h\to2a\to2b2\mu)/ \Br(h\to2a)$ as a function of $\tan \beta$ for a CP-odd scalar $a$, while in the {\it right} plot we consider a CP-even scalar $s$ for two choices of $\alpha'$ (the scalar mass is set to 40 GeV).
In both cases, the maximum value of $\Br(h\to2a(s)\to2b2\mu)/ \Br(h\to2a(s))$ of the type-III 2HDM+S ($\simeq 0.0016$) is about four times greater than that for type-I or II ($\simeq 0.0004$).

The maximum value of $\Br(h\to2a\to2b2\mu)/ \Br(h\to2a)$ in the type-III 2HDM+S model can be understood simply as follows.  
From \tabref{2HDMcoupling}, the coupling of $a b \bar{b}$ ($a \tau^+ \tau^-$ and $a \mu^+ \mu^-$) scales as $1/\tan\beta$ ($\tan\beta$).  
Thus, keeping only the most important terms and ignoring phase space and QCD corrections, 
\beq
\frac{\Br(h\to2a\to2b2\mu)}{\Br(h\to 2a)} \sim \frac{6 m_b^2 m_\mu^2}{m_\tau^4 \tan^4\!\beta+9m_b^4 \cot^4\!\beta + 6m_b^2 m_\tau^2}\,. 
\eeq
This is maximized for $\tan\beta \sim (\sqrt{3}\,m_b/m_\tau)^{1/2} \sim 2$, with the maximum value given by 
\beq
\frac{\Br(h\to2a\to2b2\mu)}{\Br(h\to 2a)} \simeq  \frac{\epsilon_{\mu\tau}}{2}, 
\eeq
where
\beq
\epsilon_{\mu\tau}\equiv \frac{\Br(a\to 2\mu)}{\Br(a \to 2\tau)}\approx \frac{m_{\mu}^2}{m_\tau^2}\approx0.0035.
\eeq
(The derivation for the CP-even scalar is identical, up to the replacement $\tan \beta \to - \sin \alpha/\cos \beta$.) Interestingly, as we discuss in \secref{discussion}, the sensitivity of a $2b2\mu$ search to $\mathrm{Br}(h\to2a)$ in these somewhat leptophilic scenarios is competitive with purely leptonic searches like $h\to2\tau2\mu$, while providing a potentially cleaner final state for experimental reconstruction. 

%%%%
\subsection{NMSSM}

An important example of a model with a non-minimal scalar sector is the NMSSM (see, \eg, \cite{Ellwanger:2009dp} for review). An additional Higgs singlet superfield
$\hat S$ is introduced to address the $\mu$ problem of the MSSM. The Higgs superpotential is given by
\beq
\mathcal W_{\text {Higgs}} \supset \lambda \hat S \hat H_u \cdot \hat H_d + \frac{\kappa}{3} \hat S^3,
\eeq
which together with the soft supersymmetry breaking terms results in the Higgs potential 
\bea
V_{\text{soft}}&\supset& m^2_{H_d} |H_d|^2+ m^2_{H_u} |H_u|^2 + m^2_S |S|^2\nonumber\\
&&+\left(\lambda A_\lambda H_u \cdot H_d S +\frac{1}{3} A_\kappa \kappa S^3 +\text{h.c.}\right),
\eea
where $\hat H_u$ and $\hat H_d$ are MSSM Higgs doublet superfields (unhatted fields indicate   complex scalar components of the hatted superfields). The parameters $\lambda$ and $\kappa$ are Yukawa couplings, while $A_\lambda$ and $A_\kappa$ are soft-breaking $A$-parameters. The resulting neutral Higgs sector contains three CP-even scalars ($h_1, h_2, h_3$) and two CP-odd ones ($a_1, a_2$), labelled in order of increasing mass.  Its phenomenology, in the context of exotic Higgs decays, can be seen as a type-II 2HDM+S model with restricted parameter choices. 

A light CP-odd scalar can be realized in the NMSSM  by taking the $R$-symmetry limit ($A_\lambda, A_\kappa\to0$)~\cite{Dobrescu:2000yn, Dermisek:2006wr, Morrissey:2008gm} or the Peccei-Quinn-symmetry limit ($\kappa, A_\kappa\to 0$)~\cite{Peccei:1977hh, Peccei:1977ur, Chun:1994ct, Hall:2004qd}.  A light CP-even or odd scalar can also occur via an accidental cancellation among parameters that control their mass.  Parameter scans have been conducted to search for NMSSM scenarios with a SM-like $\sim125 \gev$ Higgs as well as light scalars with $m_a < m_h/2$~\cite{Cerdeno:2013cz, Christensen:2013dra, Cao:2013gba, Kozaczuk:2013spa, Ellwanger:2014dfa, Bomark:2014gya}.  
If the $a$ is light, current LHC Higgs data favors it to be singlet-dominated, but $\Br(h\to 2a)\sim \OO(10\%)$ is possible in the surviving parameter space. 

It is interesting to consider the possible connection between $h\to2a$ decays and naturalness in NMSSM models. An NMSSM scenario can be considered potentially natural if radiative Higgs mass corrections are small compared to tree-level contributions. 

If $h = h_1$ and $a=a_1$, the tree-level SM Higgs mass is given by
\bea
m^2_{h, \text{tree}} &\simeq& m_Z^2 \cos^2 2\beta +\lambda^2 v^2 \sin^2 2 \beta\nonumber\\
&& -\frac{\lambda^2 v^2}{\kappa^2}\left[\lambda-\sin 2\beta \left(\kappa+\frac{A_\lambda}{2s}\right)\right]^2,
\eea
where $s\equiv\vev S$ and $\tan \beta \equiv \vev {H_u}/\vev{H_d}$. As argued in~\cite{Ellwanger:2009dp, Hall:2011aa}, the naturalness limit of the NMSSM is reached for low $\tan \beta$ and $\lambda$ as large as possible (perturbativity at the GUT scale bounds $\lambda \lesssim 0.7$). Since the triple Higgs coupling $h_1a_1a_1$ is proportional to $\lambda$ at tree-level in 
the NMSSM, $\lambda\approx 0.7$ would imply $\Br(h_1 \to 2 a_1)\approx 100\%$ if the channel is kinematically accessible, which is strongly disfavored by current LHC data.  
Therefore, the surviving parameter space with a sufficiently small $\Br(h_1\to 2 a_1)\lesssim 0.1$ requires a somewhat unnatural 
realization of the NMSSM in this scenario.

For $h = h_2$ and $a=a_1$, mixing in the CP-even scalar sector can help to increase $m_{h_2}$~\cite{Badziak:2013bda}. 
The naturalness limit with $m_{a_1}<m_{h_2}/2$ is accommodated with $\tan \beta\sim 4-6$ and the 
comparatively smaller $\lambda \lesssim 0.4-0.5$~\cite{Bomark:2014gya}.  This allows for $\Br(h_2\to 2a_1)\lesssim 0.1$, 
consistent with current LHC data.  
This conclusion is supported by~\cite{Cao:2013gba}.

%%%%%%%%%%%%%%%%%%%%%%%%%%%%%%%%%%%%%%%%%%%%%%%%%%%%%%%%%%%%%%%%%%%%%%
%%%%%%%%%%%%%%%%%%%%%%%%%%%%%%%%%%%%%%%%%%%%%%%%%%%%%%%%%%%%%%%%%%%%%
\section{Reach Estimate}
\label{sec:ConstraingExclusionyPotential}

In this section, we estimate the reach of the search for $h\to 2a \to 2b2\mu$ with $20 \fb$ at the 8~TeV LHC, and with $30\fb$, $300\fb$, and $3000\fb$ at the 14~TeV LHC.  For simplicity, we only consider $a$ to be a CP-odd scalar and the two intermediate $a$'s to be identical and on-shell.  These results should apply, with  little modification, to the case where the intermediate state is CP-even, as 
we do not make explicit use of any angular information of the decay.

We assume that the 125 GeV Higgs boson, $h$, is SM-like except for a non-zero branching ratio for the exotic decay $h \to2a$.  In particular, we assume that $h$ is mainly produced through ggF and has a non-zero branching ratio for the decay $h \to2a\to2b2\mu$. Higgs production via vector boson fusion is not included in our analysis, making our projected sensitivities slightly pessimistic. The signal is simulated for the mass of $a$ ranging from 15~GeV to 60~GeV. Lower masses of $a$ (but still above the $2b$ threshold) may involve complicated decays to bottomonium and are beyond the scope of this study~\cite{Baumgart:2012pj}. 

We will consider three types of analyses below.  A ``conventional analysis'' (\S\ref{sec:ExpectedSensitivity}) 
will make use of standard anti-$k_t$ jets (from $a\to 2b$) with a radius of $R=0.4$ or $R=0.5$.  
For low $m_a$, these jets are boosted and merge, so that an analysis with $R=0.2$ is more sensitive (\S\ref{sec:SmallRadius}).  
Finally, we use jet-substructure techniques to improve the low-$m_a$ reach further (\S\ref{sec:JetSubstructures}).  

The dominant backgrounds are Drell-Yan (DY) production with associated jets, \ie, $Z^{(*)}/\gamma^*+2b/2c/2j$, where $Z^{(*)}/\gamma^*$ produces a muon pair.\footnote{We have checked that the corresponding background where $Z^{(*)}/\gamma$ produces two leptonic $\tau$'s is negligible in our analysis, due to the larger amount of missing energy and our strict $m_h$ reconstruction requirement.} A secondary background arises from $t\bar t$ production. Backgrounds from diboson production ($ZZ$, $WW$, $WZ$) have small enough cross sections so that we can neglect them. 
Finally, it is possible for QCD multi-jet events, with two jets being mis-identified as muons, to contribute to the background.  These `lepton fakes' are notoriously difficult to simulate. In Appendix~\ref{sec:MultiJet}, we use the methods of~\cite{Curtin:2013zua} to estimate their importance compared to the irreducible DY backgrounds. We find that it is reasonable to neglect muon fakes for an analysis with 0 or 2 $b$-tags, but they may be competitive if we require only a single $b$-tag.  We therefore limit ourselves to 
using either $0$ or $2$ $b$-tags in \S\ref{sec:ExpectedSensitivity} and \S\ref{sec:SmallRadius}; in these analyses, we find in any case that the sensitivity is not noticeably improved by including a single $b$-tag.  However, in \S\ref{sec:JetSubstructures}, we 
consider the possibility of requiring a single fat jet with a single $b$-tag.  For this, a data-driven estimate of lepton-fakes 
to determine their importance will be needed by the experimental collaborations.

%%%%%%%%%%%%%%%%%%%%%%%%%%%%%%%%%%%%%%%%%%%%%%%%%%%%%%%%%%%%%%%%%%%%%%
\subsection{Conventional analysis}
\label{sec:ExpectedSensitivity}

\begin{table}[t!]
\begin{tabular}{|c|c|c|}
\hline
 & 8 TeV cross section (pb) & 14 TeV cross section (pb)\\
 \hline
$b \bar b \mu^+ \mu^-$        & 6.11                      & 12.16                      \\
$c \bar c \mu^+ \mu^-$        & 60.44                     & 109.50                     \\
$jj \mu^+ \mu^-$        & 151.65                    & 275.17                     \\
$t\bar t$   & 0.68                      & 2.49			\\
\hline
\hline
$ jj \mu^+ \mu^{-*}$        & 152.24                    & 279.17                     \\
\hline
\end{tabular}
   \caption{Cross sections for various backgrounds after applying generator level cuts as described in \S\ref{sec:ExpectedSensitivity}, given by \texttt{Sherpa 2.1.1}. The last row refers to DY $Z^{(*)}/\gamma^*+2j$ background with different generator level cuts, as required for the small-radius jets and jet substructure analyses in \S\ref{sec:SmallRadius} and \S\ref{sec:JetSubstructures}. 
   These cross sections are scaled in our reach estimates by a pessimistic $K$-factor of 2 to account for higher-order effects.}
   \label{tab:background_xsec}
\end{table}

Signal, as well as DY $Z^{(*)}/\gamma^*+2b/2c/2j$ and $t\bar t$ backgrounds, are simulated at leading-order (LO) by \texttt{Sherpa 2.1.1}~\cite{Gleisberg:2008ta} for the 8 and 14 TeV LHC with the CT10~\cite{Lai:2010vv}
parton distribution function (PDF), and matched up to three jets 
(i.e., for example, we include one extra jet for the signal).  
We ignore lepton fakes from pure QCD, as justified in Appendix~\ref{sec:MultiJet}.
At generator level, no cut is imposed on the signal. The generator-level cuts for the backgrounds are: $p_{T\,\mu} > 5 \gev$, $|\eta_\mu|<5$ and $10 \gev < m_{\mu \mu} < 70 \gev$. Additionally, for $Z^{(*)}/\gamma^*+2j$ we require at least two partons with $p_{T\, j} > 10\gev$ and $|\eta_j| < 5$. Here, $j$ refers to partons clustered into jets with the anti-$k_t$ algorithm with radius $R = 0.2$.

The signal cross sections are normalized to $\sigma_{ggF} \times \mathrm{Br}(h \to 2b2\mu)$, where 
$\sigma_{ggF}\simeq 19.3 \pb$ and $49.47 \pb$ are the next-to-leading-order (NLO) ggF Higgs production cross section for 8~TeV and 14~TeV, respectively~\cite{Dittmaier:2011ti}. Given the generator level cuts as described above, the cross sections for the backgrounds given by \texttt{Sherpa 2.1.1} are shown in \tabref{background_xsec}. We then scale all  backgrounds by a pessimistic 
$K$-factor of 2, to account for higher-order effects in our sensitivity estimates.

\begin{table}[t!]
\renewcommand{\arraystretch}{1.6}
\begin{tabular}{|c|c|c|}
\hline
 & ATLAS & CMS 
 \\ \hline \hline
$\epsilon_\mu$ 
& $\displaystyle \left\{ \begin{array}{ll} 0.98 & p_T > 6 \gev \\ 0 & \mathrm{otherwise} \end{array} \right.$ 
& 
$\displaystyle \left\{ \begin{array}{ll} 0.96 & p_T > 6 \gev \\ 0 & \mathrm{otherwise} \end{array} \right.$
\\ \hline
$\Delta R^\mathrm{iso}_{\mu}$ & 0.3 & 0.4 
\\ \hline
$\mathrm{max} \left(p_T^\mathrm{cone}/p_T^\mathrm{\mu}\right)$
& 1.13
& 1.10
\\ \hline \hline
$R_\mathrm{jet}$ & 0.4 & 0.5
\\ \hline
$R_\mathrm{microjet}$ & 0.2 & 0.2
\\ \hline 
\end{tabular}
\caption{
Relevant object reconstruction parameters assumed for the ATLAS and CMS detectors. $\epsilon_\mu$ is the muon tagging efficiency for $|\eta| < 2.4$
(Note that our analysis relies on a dimuon trigger, which has a higher threshold than 6 GeV.)
For a muon with $p_T^\mu$ to pass the isolation criteria, the $p_T$ of all the objects in a cone of radius $\Delta R^\mathrm{iso}_\mu$  around the muon must be less than the shown $\mathrm{max}\left(p_T^\mathrm{cone}/p_T^\mathrm{\mu}\right)$. Jets are anti-$k_T$ clustered~\cite{Cacciari:2008gp} with a radius given by $R_\mathrm{jet}$. For the analysis in \S\ref{sec:SmallRadius}, 
this is reduced to 0.2. See text for details on $b$-tagging.
}
\label{tab:detector}
\end{table}

%%%%%%%%%%%%%%%%%%%%%%%%
\begin{table*}[htbp]
{%
   \centering
   \begin{tabular}{@{} |l|c|cccc| @{}} 
      \hline

      Selection Criteria    & $S$ & $b \bar b \mu^+ \mu^-$ & $c \bar c \mu^+ \mu^-$ & $jj\mu^+ \mu^-$ & $t\bar t$\\
\hline
\hline
 && \multicolumn{4}{|c|}{ $p_{T\,\mu} > 5 \gev$, $|\eta_\mu|<5$ and $10 \gev < m_{\mu \mu} < 70 \gev$}   \\
Generator level cuts & no cuts & \multicolumn{4}{|c|}{ for $jj\mu^+ \mu^-$, we require  in addition two partons }  \\
 && \multicolumn{4}{|c|}{with $p_{T\, j} > 10\gev$ and $|\eta_j| < 5$}\\
\hline
 $N_{\rm ev,~gen.}$ ($20\fb$) & $6.3\times10^2$ & $2.4\times 10^5$  &$2.4\times 10^6$& $6.1\times 10^{6}$ &$2.7\times 10^4$ \\
 \hline
\hline
  pass OS dimuon trigger      &   & &  &  & \\
$p_{T\, \mu_1,\mu_2}\! >\! (13,13) \gev$ or $(18, 8)  \gev$     &  50\%  & 27\% &  19\% & 29\% & 60\% \\

       at least  two $b$-jets with      &&&&&    \\
       $p_{T\,b}>25\gev$ and $|\eta_b|<2.5$    & 3.8\%  &  17\% & 1.3\% & 0.45\% & 37\%  \\
  $\Delta R_{b_1 b_2, b\mu, \mu_1\mu_2}>0.4, 0.4, 0.3$ & 99\%& 99\%&  99\%& 99\%&99\%\\ 
      \hline
       $N_{\rm ev, presel.}$ ($20\fb$) & $12$ & $1.1\times 10^4$  &$5.7\times 10^3$& $7.9\times 10^3$ &$5.9\times 10^3$ \\
      \hline
      \hline

$\slashed{E}_T< 30~\text{GeV}$ & 98\% & 90\% & 95\% & 92\%& 12\%  \\

     $|m_{b_1b_2\mu_1\mu_2}-m_h|< 15$ GeV& 54\% & 4.7\% & 3.3\% & 3.3\% & 0.6\%   \\

$|m_{b_1b_2}-m_a| < 15~\text{GeV}$ & 97\% & 25\% & 31\% & 61\% & 24\%  \\

    $|m_{\mu_1\mu_2}-m_a|< 1$ GeV      &  100\%  & 3.4\% &  2.9\% & 3.7\% & 7.6\%  \\
   \hline

$N_{\rm ev,~final}$ ($20 \fb$) & 6.2 & 4.0  & 1.6 & 5.3 & $0.1$ \\
 \hline
   \hline

   \multicolumn{6}{|c|}{$S=6.2$,\quad$B_{\text{tot}}=11$,\quad$S/B_{\text{tot}}=0.6$, \quad $S/\sqrt{B_{\text{tot}}}=1.9$}\\

      \hline
   \end{tabular}}
   \caption{Relative efficiencies for the signal ($S$) $h\to aa \to b \bar b \mu^+ \mu^-$ ($m_a = 40 \gev$) and indicated backgrounds, with 2 $b$-tags at ATLAS 8 TeV. All signals and backgrounds listed are simulated with \texttt{Sherpa 2.1.1}. The number of signal and background events after passing the generator level cuts, preselection cuts, and higher level cuts are also listed as $N_{\text{ev, gen.}}$, $N_{\rm ev, presel.}$, and $N_{\rm ev, final}$, respectively. (Meaningful comparisons are only possible between the latter two as $N_{\text{ev, gen.}}$ is biased by different generator-level cuts on signal and background.)
    For the signal normalization, we take the NLO ggF production cross section $\sigma_{ggF} = 19.3 \pb$~\cite{Dittmaier:2011ti}, and assume $\Br(h\to aa) \approx 100\%$, $2\times \Br (a\to b \bar b) \Br (a\to \mu^+ \mu^-)=1.6\times 10^{-3}$. The latter branching ratio factor corresponds to a 2HDM model of type-III plus a singlet with $\tan \beta=2$ (see \S\ref{2HDMS} and \figref{Br-vs-tanbeta} ). 
For the background normalization, we adopted cross sections at generator level from \texttt{Sherpa} (see \tabref{background_xsec}) and scaled them by a pessimistic $K$-factor of 2.}
   \label{tab:2b2mu_cuts}
   \end{table*}   

   \begin{figure*}[htbp]
   \centering
   \includegraphics[width=0.32\textwidth]{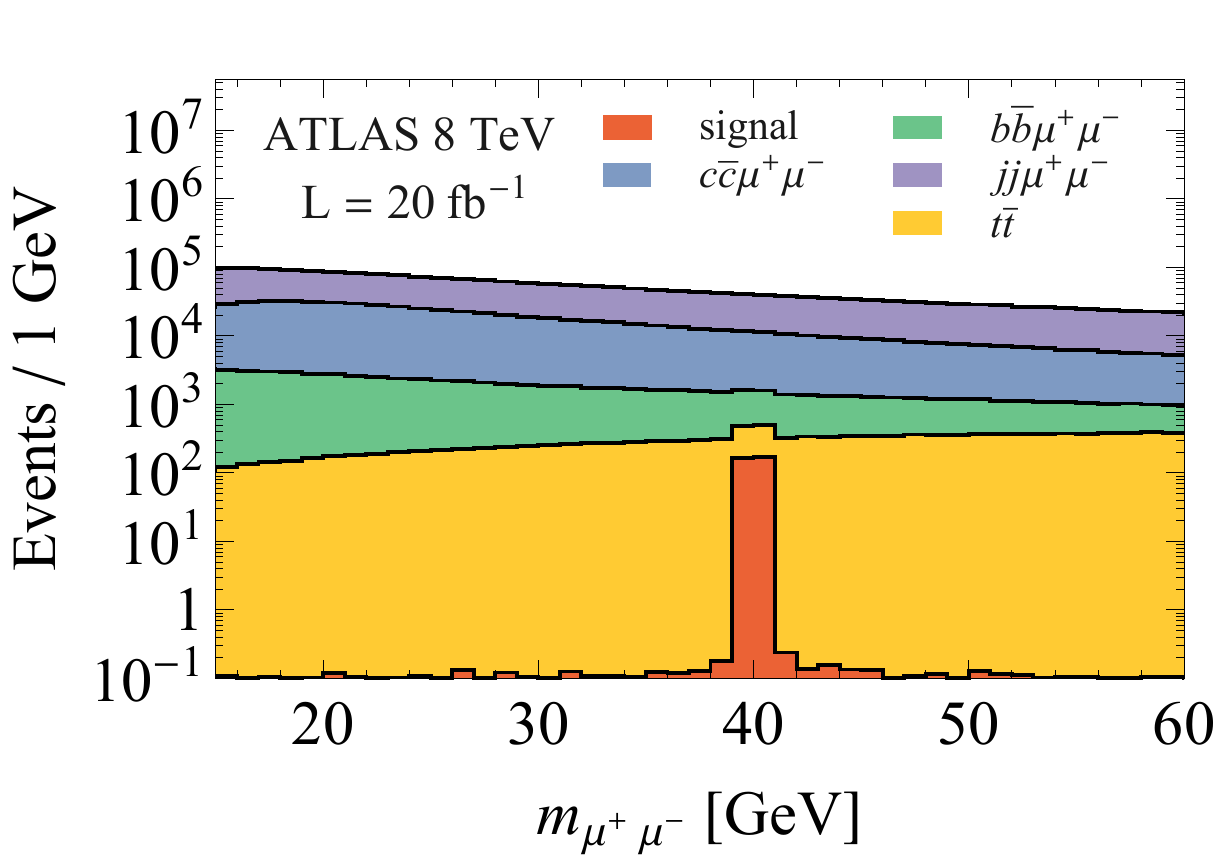}
   ~\includegraphics[width=0.32 \textwidth]{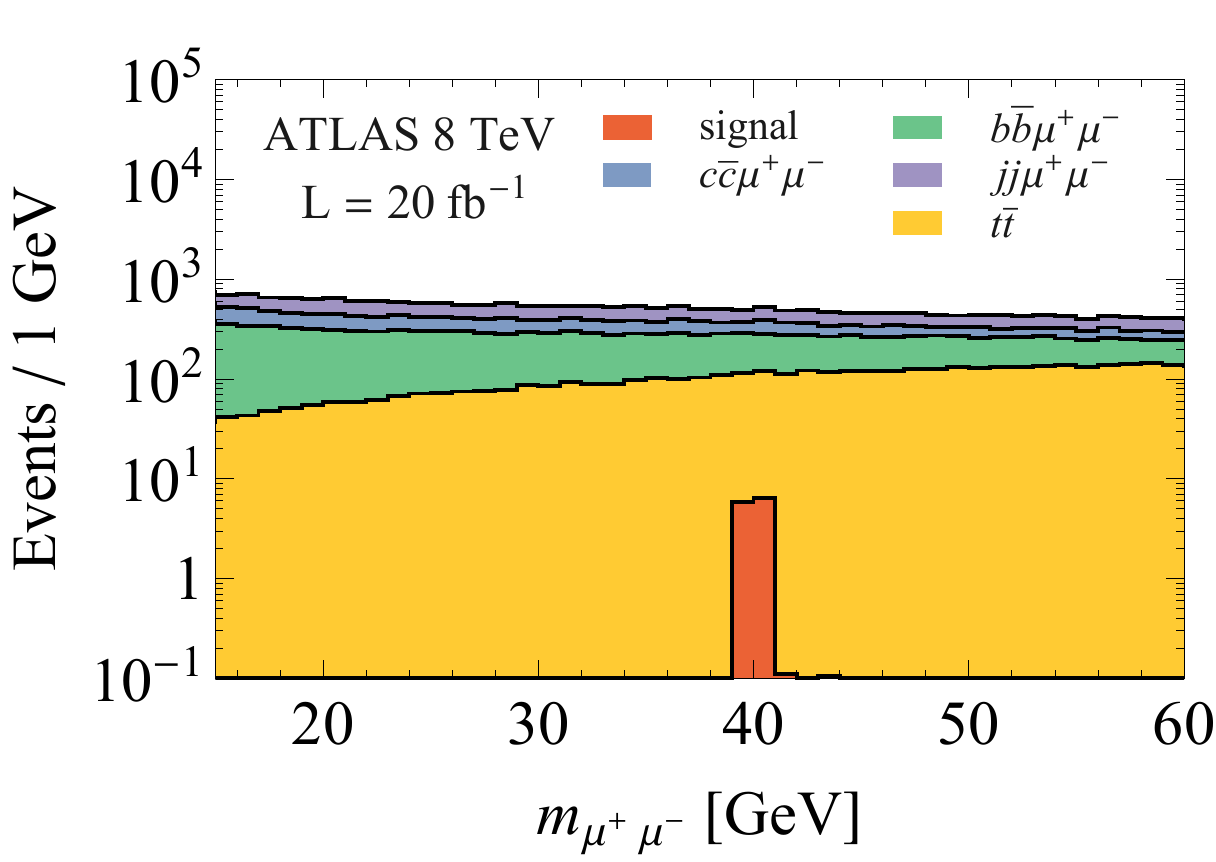}      
    ~\includegraphics[width=0.305 \textwidth]{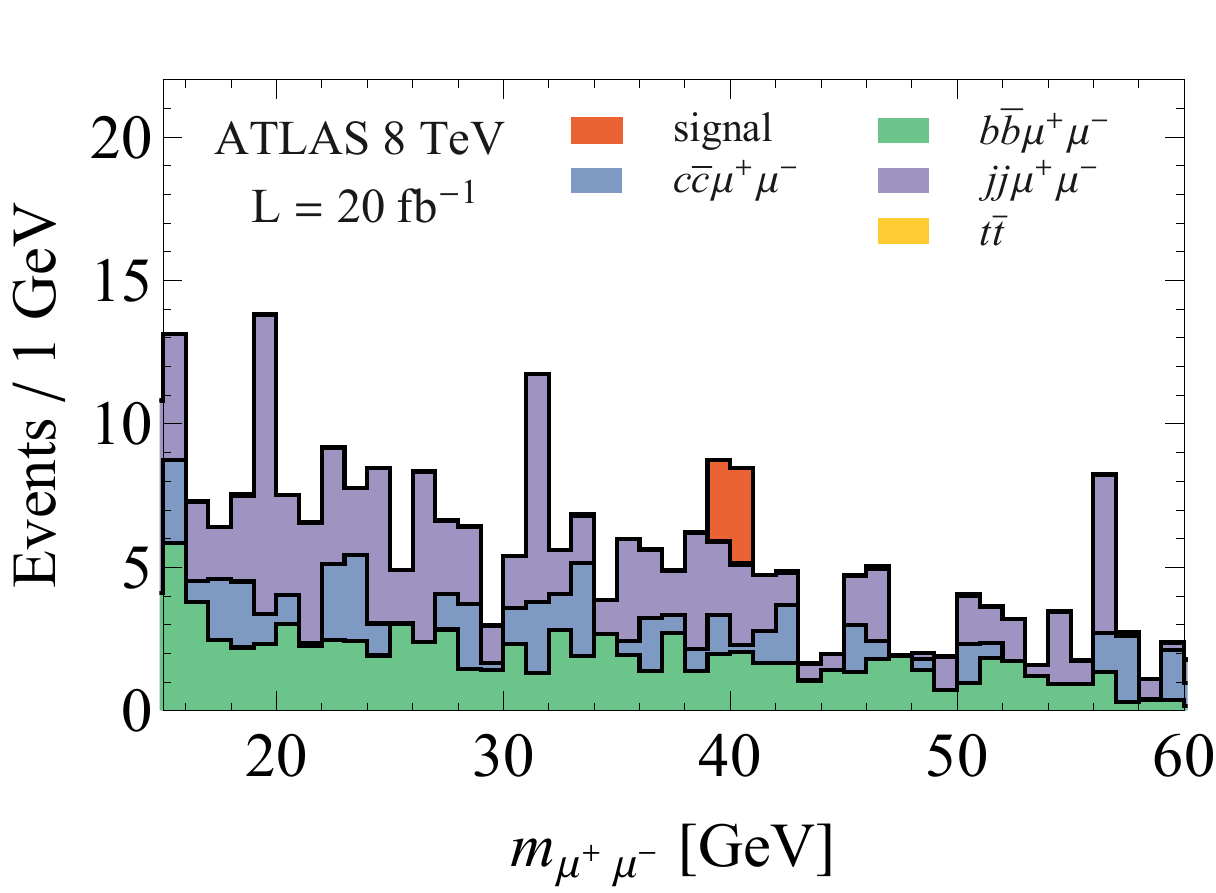} 
      \caption{Stacked $m_{\mu^+\mu^-}$ distributions for signal ($m_a= 40 \gev$) and backgrounds with 2 $b$-tags at ATLAS 8 TeV for $20 \fb$. The \emph{left}, \emph{center}, and \emph{right} plots represent the distributions after passing the generator level cuts, preselection cuts, and higher level cuts respectively. In the {\it right} plot, all cuts have been included {\it except} for the cut on $m_{\mu_1\mu_2}$.  
      We choose the signal size to correspond to $\sim 2\sigma$ sensitivity of our analysis.  The assumptions for  cross sections and branching ratios are the same as in \tabref{2b2mu_cuts}.}
   \label{fig:2b2mu_histogram}
\end{figure*}
\begin{figure*}[htbp]
   \centering
   \includegraphics[width=0.48\textwidth]{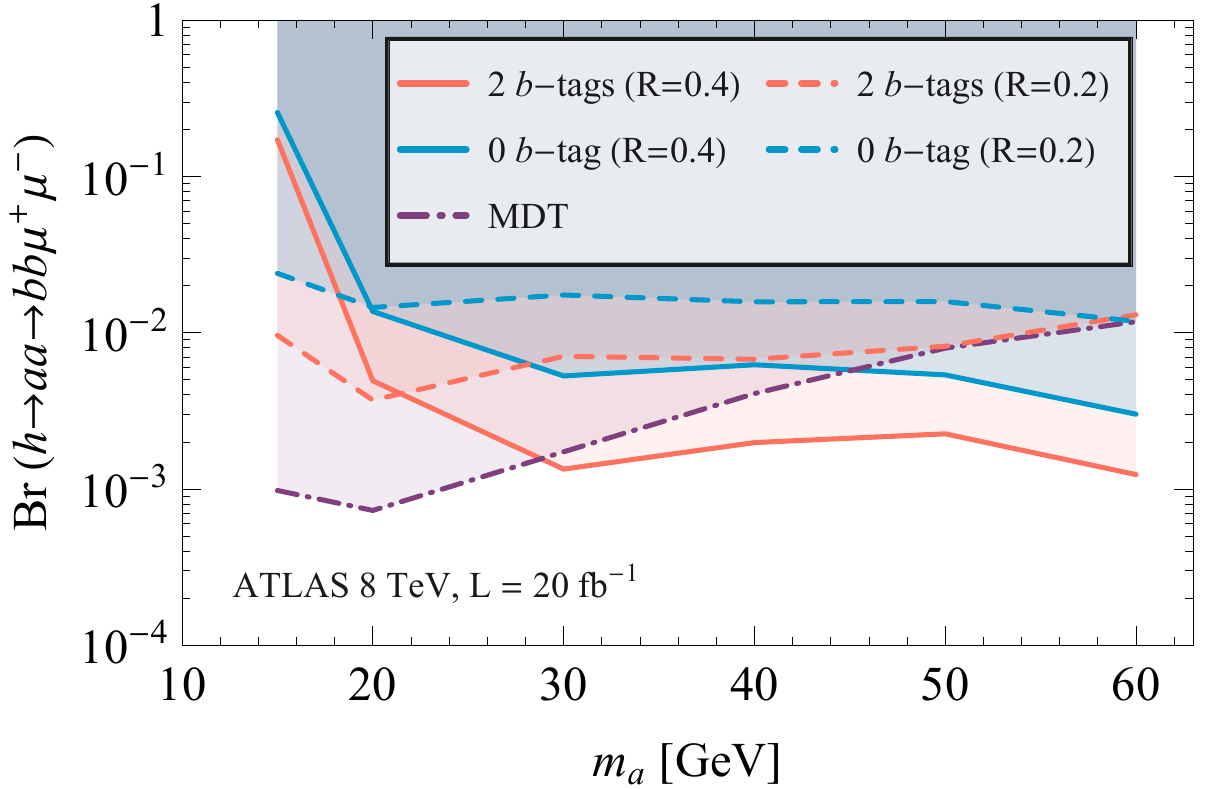}
   ~\includegraphics[width=0.48 \textwidth]{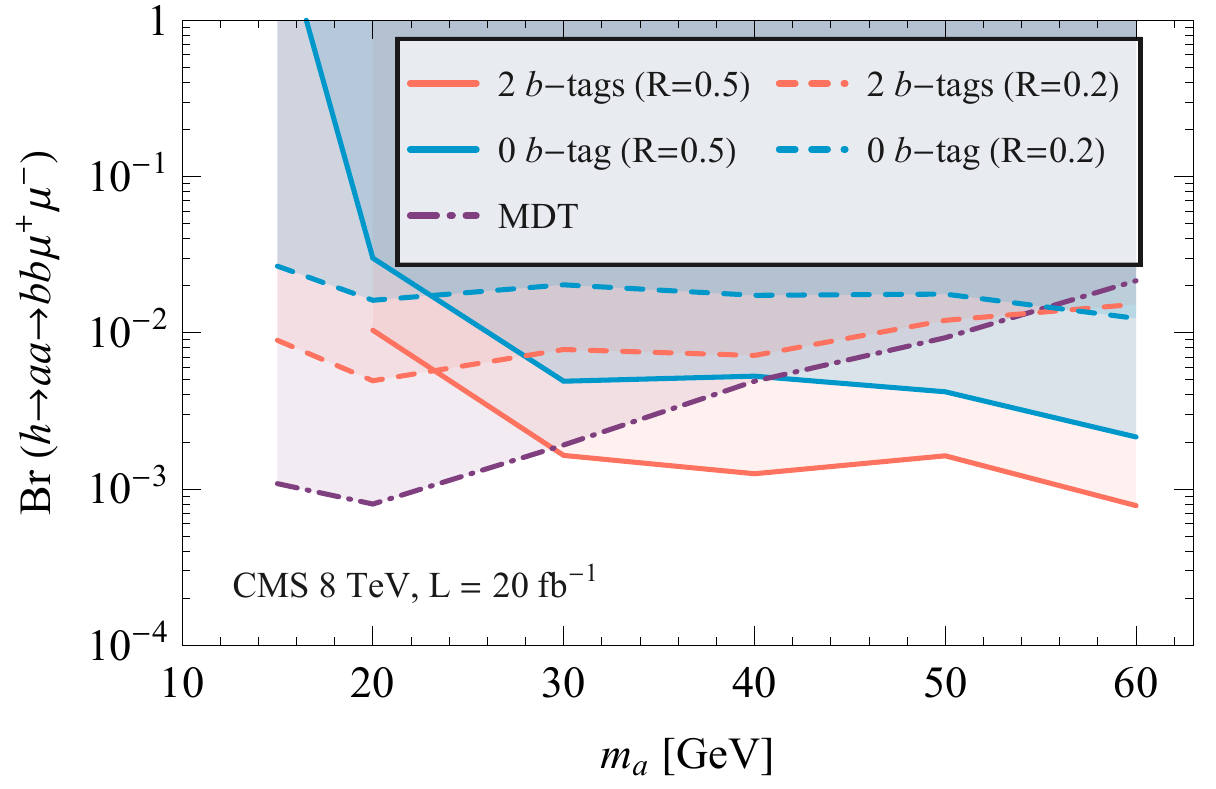}      
      \caption{Expected 95\% CL sensitivity to Br$(h\to aa \to b \bar b \mu^+ \mu^-)$ for 20 $\fb$ data at  8 TeV ATLAS (\emph{left}) and CMS (\emph{right}). The solid line is the sensitivity of the ``conventional'' analysis (\secref{ExpectedSensitivity}) with 
      a jet-clustering radius of either $R = 0.4$ (ATLAS) or $0.5$ (CMS).  The sensitivities when using a smaller jet radius of $R = 0.2$ (\secref{SmallRadius}) is shown with dashed lines. The purple dot-dashed line is the sensitivity from a jet substructure analysis that makes use of the mass drop tagger (MDT) (\secref{JetSubstructures}).  
      }
   \label{fig:8tev}
\end{figure*}

Detector simulation and data analysis are performed by an in-house software framework also used e.g.~in~\cite{Curtin:2012rm, Curtin:2013zua, Curtin:2014zua}.  This includes jet clustering with \texttt{FastJet 3.0.6}~\cite{Cacciari:2011ma}, application of realistic efficiency curves and isolation requirements for $b$-jet and lepton reconstruction, and geometric detector acceptances. The relevant detector parameters for our analysis are given in \tabref{detector}. The differences between the two detectors' capabilities are relatively minor and the projected limits for both will be similar. However, the larger jet clustering radius in the CMS conventional analysis will affect the low-mass limit. 
 We adopt the $b$-tagging efficiency curve for the ``MV1'' algorithm at the 70\% $b$-jet efficiency working point in~\cite{ATLAS:2014cal} and the $c$/light-jet rejection curves with respect to $b$-jet efficiency (also for the MV1 algorithm) in~\cite{ATLAS:2012ima}. 
For jet $p_{T}$ of around 200 GeV, the $b$-tagging efficiencies for $(b,c,\mathrm{light})$ jets are $(0.78, 0.3, 0.03)$. 
These efficiencies drop to $(0.54, 0.1, 0.001)$ at $p_T = 25 \gev$.  
We use the same $b$-tagging efficiencies for both the ATLAS and CMS analyses.

The events will be recorded using a di-muon trigger.  For the LHC 8 TeV search, we impose the dimuon trigger used in~\cite{Aad:2014fia}, requiring $|\eta_{\mu_1, \mu_2}|<2.4$ and $p_{T\, \mu_1,\mu_2}\! >\! 13 \gev, 13 \gev$ or $p_{T\, \mu_1,\mu_2}\! >\! 18 \gev, 8\gev$ (objects are  labelled in order of decreasing  $p_T$). We then impose several ``preselection cuts''.  
The leading jets are required to satisfy $p_T > 25 \gev$, $|\eta| < 2.5$, and $\Delta R_{J_1 J_2} > 0.4$.  
On the two (leading) muons we impose $\Delta R_{\mu_1\mu_2} > 0.3$. The distances between the two leading  jets and the two leading  muons must satisfy $\Delta R_{J\mu} > 0.4$ ($J$ stands for the two leading jets ($b$-jets) for the analysis with 0 (2) $b$-tags).  Events with either 0 or 2 $b$-tags are selected. 

Following this preselection, we now impose cuts to separate the signal from background. A missing transverse energy cut of $\slashed{E}_T < 30 \gev$ suppresses $t \bar t$ background. We also make use of the double-resonance structure of the signal by imposing mass reconstruction cuts
\bea
|m_{J_1 J_2 \mu_1 \mu_2}-m_h|&<& 15\gev,\nonumber\\
|m_{J_1J_2}-m_a|&<& 15\gev,\nonumber\\
|m_{\mu_1\mu_2}-m_a| &<& 1\gev,
\eea
separately for each $m_a$. 

\tabref{2b2mu_cuts} shows an example of the relative efficiencies for the signal with $m_a=40\gev$ and backgrounds with 2 $b$-tags for ATLAS at 8 TeV. \figref{2b2mu_histogram} shows the corresponding stacked histograms for the signal and backgrounds after passing the generator level, preselection level, and higher level cuts (except for the cut on $m_{\mu_1\mu_2}$). 
Despite simulating a very large number of events, our background $m_{\mu\mu}$ spectra display some fluctuations after all the other cuts with two $b$-tags are applied. This can partially be attributed to the way \texttt{Sherpa} generates weighted events, but is more generally due to the difficulty of overpopulating each small $m_{\mu\mu}$ bin in our signal region with DY+jets Monte Carlo, in order to determine the expected number of background events with high precision. 
However, at the level of precision of our study, this will not significantly affect our derived sensitivity reach, for which we assume a simple counting experiment after applying the above cuts, with the background expectation taken directly from the Monte Carlo prediction. For an experimental study, a side-band-type analysis would be used to estimate the SM contribution in a particular $m_{\mu\mu}$ bin directly from data. Since the aim of our study is merely to estimate the $2\sigma$ exclusion potential, we can neglect these details, including systematic uncertainties, which we have no way of reliably determining.  In particular, we also do not show the $5 \sigma$ discovery reach, as this would require an estimate of the look-elsewhere effect, which depends on how the analysis is done. 

The expected 95\% confidence level (CL) sensitivity to $\Br(h\to aa \to b \bar b \mu^+ \mu^-)$ from 8 TeV data are shown in \figref{8tev} for both ATLAS and CMS. Requiring 2 $b$-tags increases the sensitivity by about a factor of 3 compared to requiring no $b$-tags. The expected bounds are approximately independent of scalar mass for $m_a \geq 30 \gev$. For $m_a < 20 \gev$, the signal efficiency drops dramatically because the two $b$'s from the $a$-decay become collimated. In fact, in our CMS analysis  (which required the jets to satisfy $R=0.5$), no signal events passed the cuts for this low $m_a$ region. However, as we show in the next sections, $b$-tagging with a smaller jet radius or the use of jet substructure can dramatically  improve sensitivity in this region. 

\begin{figure*}[htbp]
   \centering
    \includegraphics[width=0.48\textwidth]{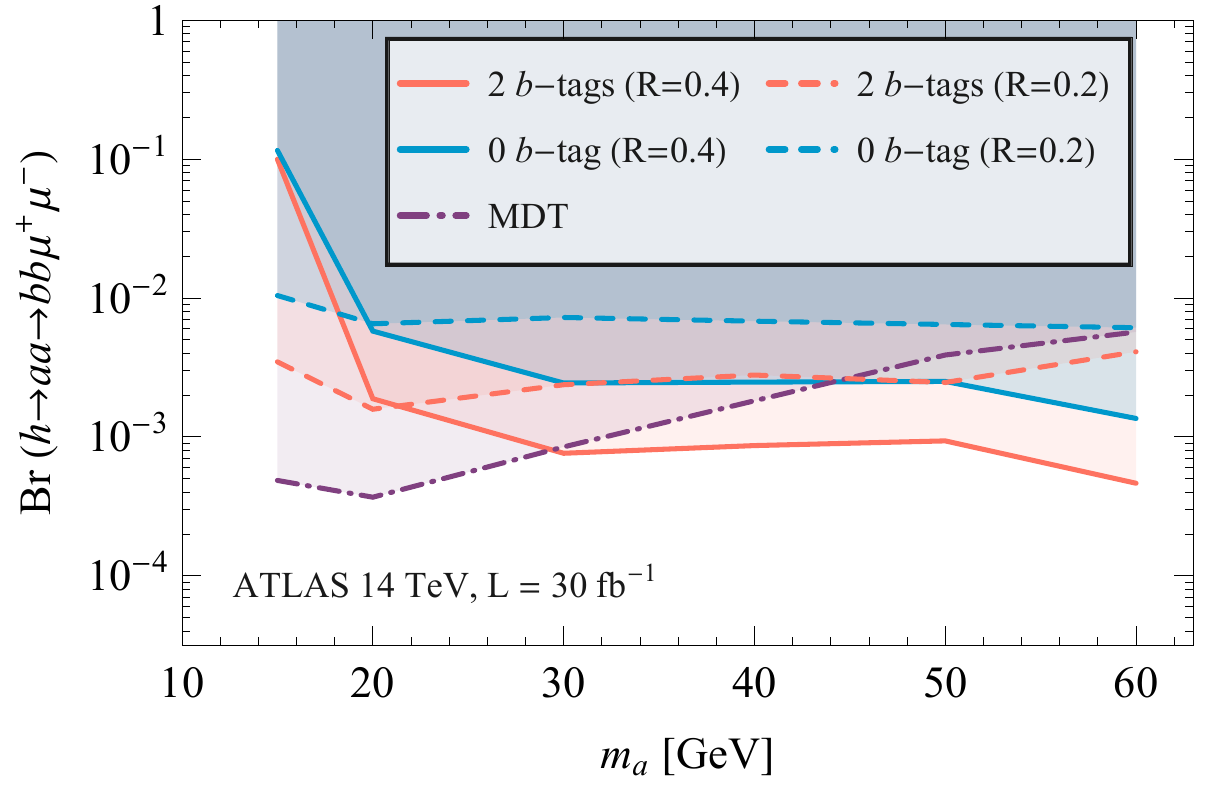}
   ~\includegraphics[width=0.48 \textwidth]{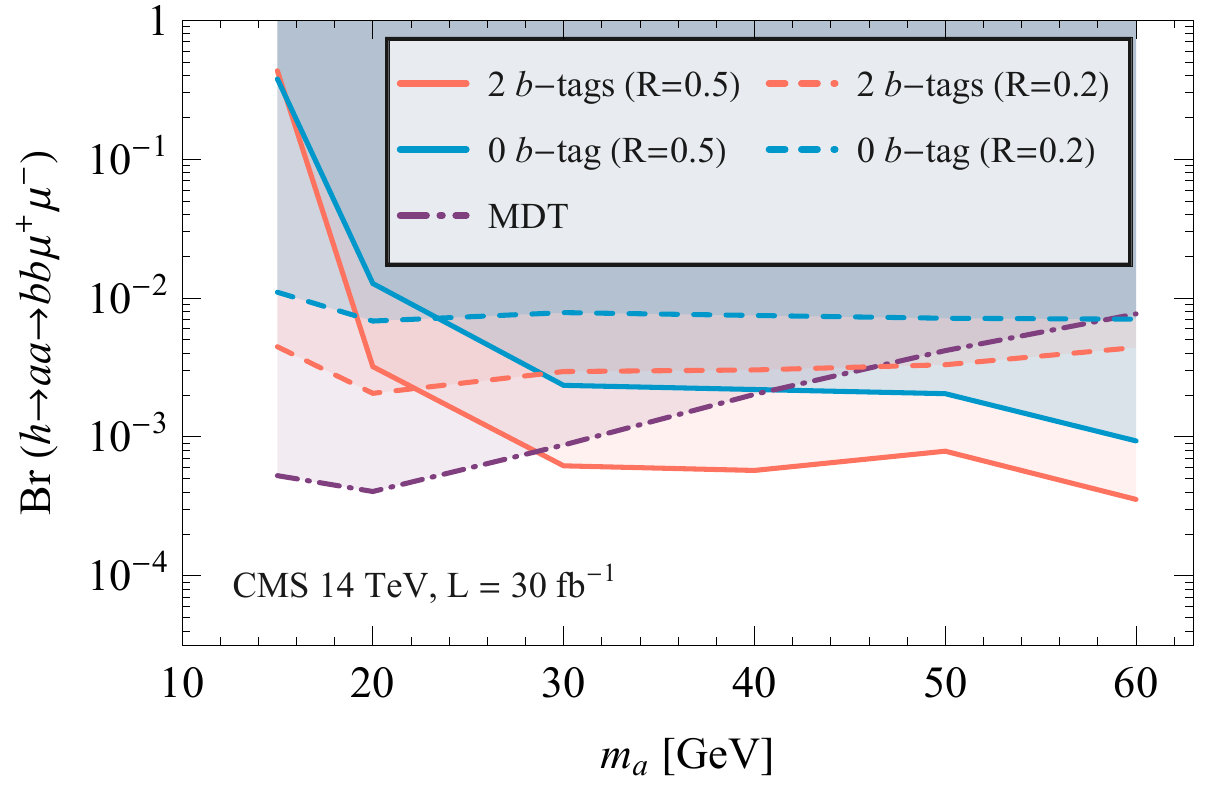} \\
   \includegraphics[width=0.48\textwidth]{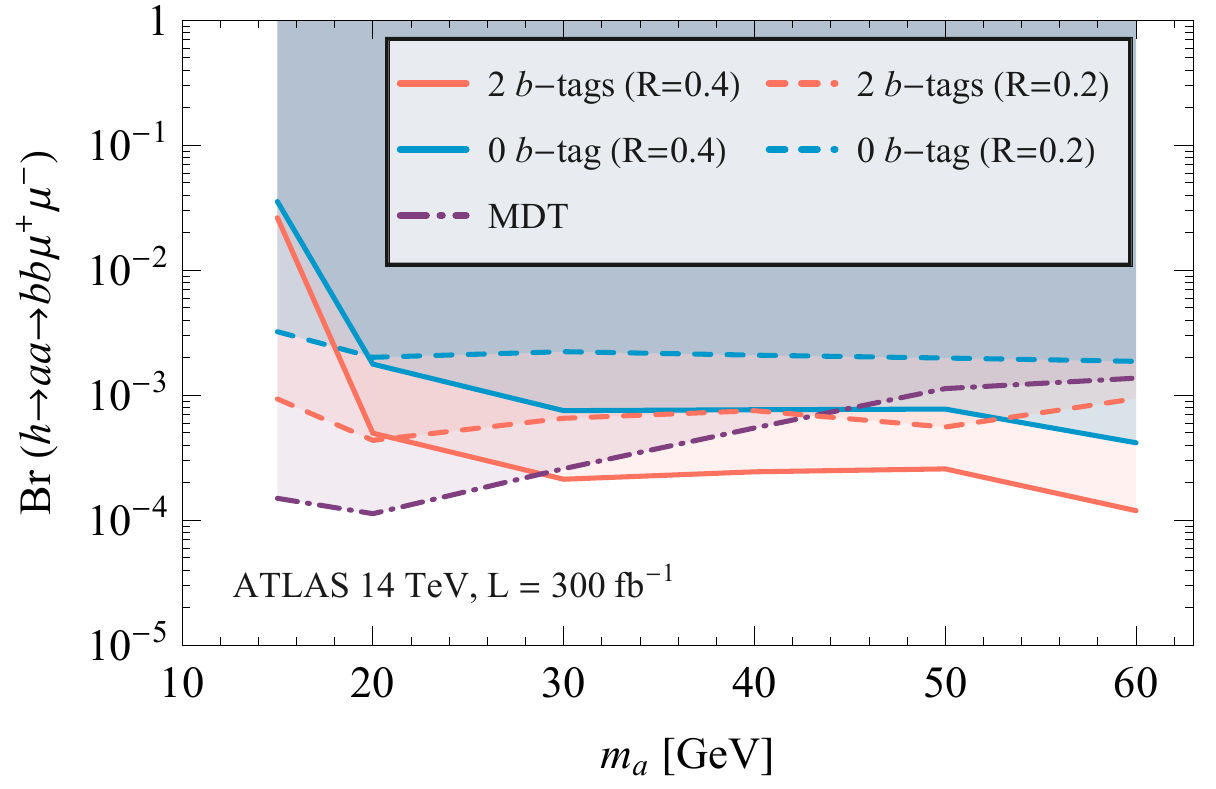}
   ~\includegraphics[width=0.48 \textwidth]{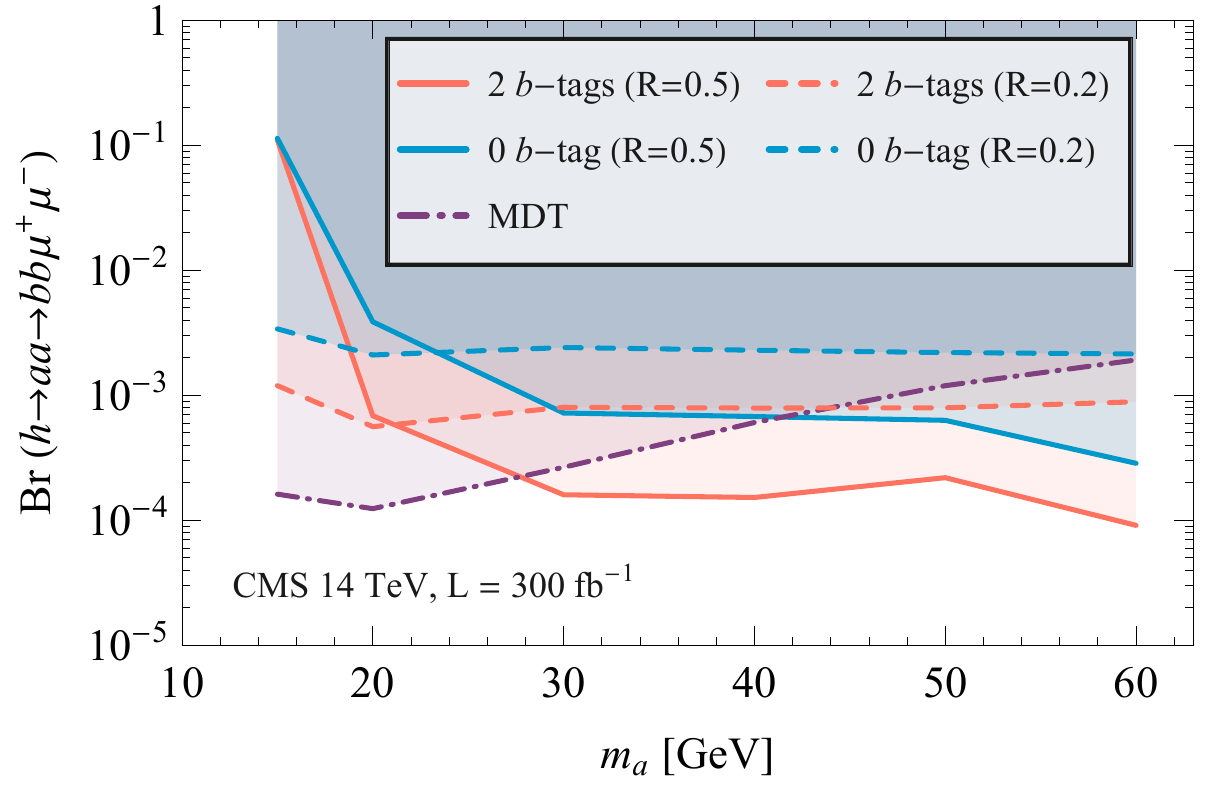} \\
        \includegraphics[width=0.48\textwidth]{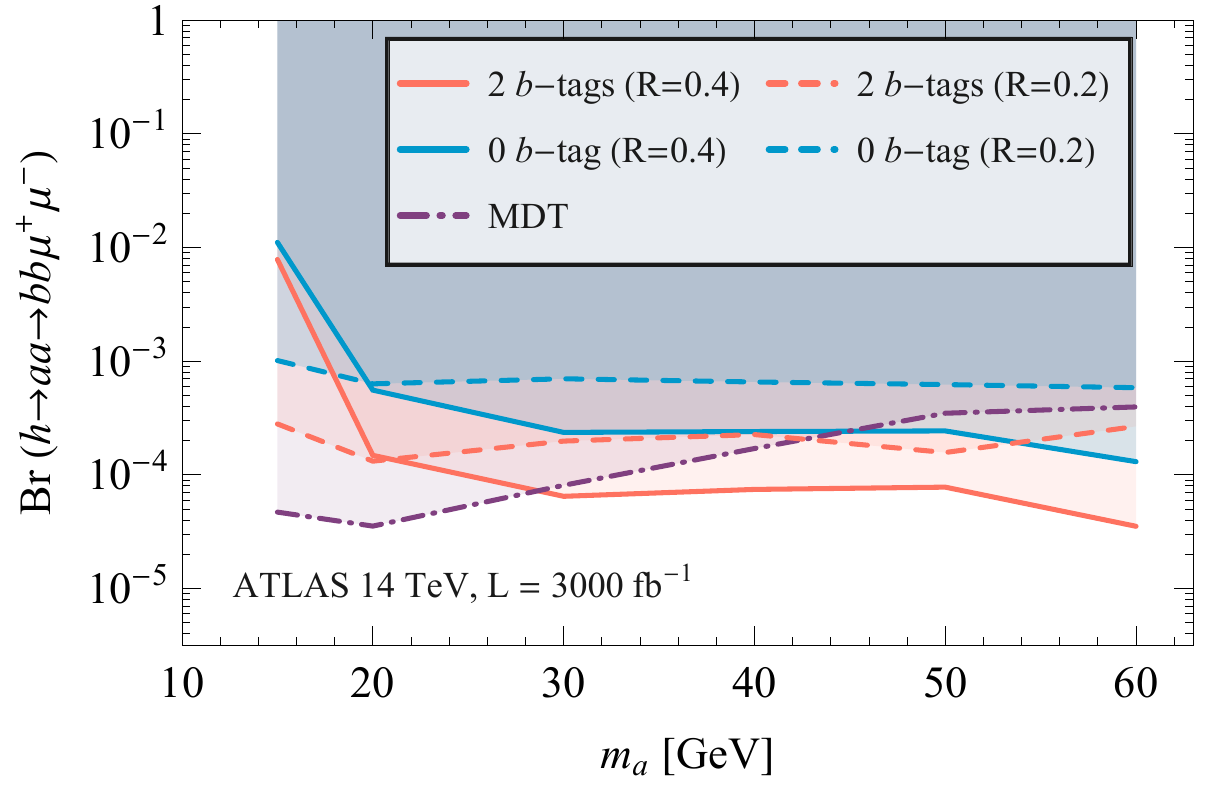}
   ~\includegraphics[width=0.48 \textwidth]{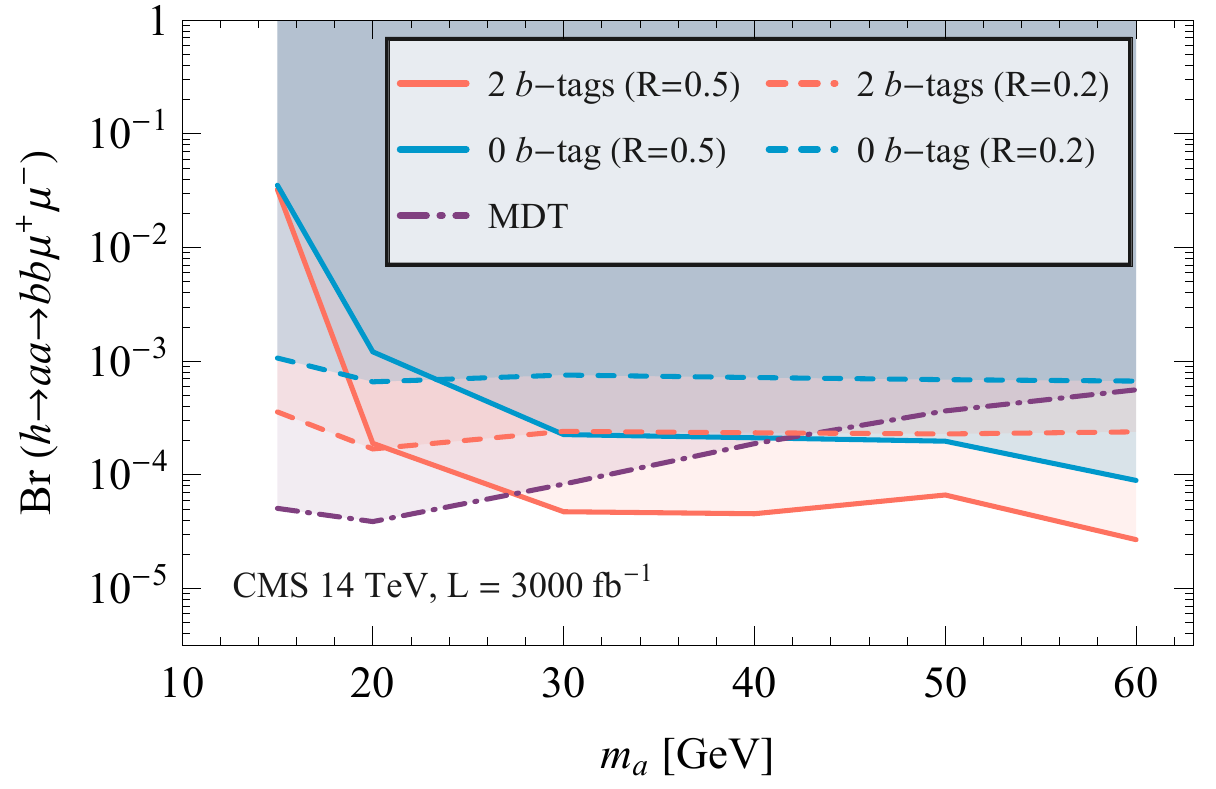} 
      \caption{Expected 95\% CL reach on Br$(h\to aa \to b \bar b \mu^+ \mu^-)$ for 30 (\emph{top}), 300 (\emph{center}), and 3000 (\emph{bottom}) $\fb$ at 14 TeV, for ATLAS (\emph{left}) and CMS (\emph{right}). 
      The solid line is the sensitivity of the ``conventional'' analysis (\secref{ExpectedSensitivity}) with 
      a jet-clustering radius of either $R = 0.4$ (ATLAS) or $0.5$ (CMS).  The sensitivities when using a smaller jet radius of $R = 0.2$ (\secref{SmallRadius}) is shown with dashed lines. The purple dot-dashed line is the sensitivity from a jet substructure analysis that makes use of the mass drop tagger (MDT) (\secref{JetSubstructures}).  
      }
   \label{fig:14tev}
\end{figure*}

The analysis proceeds nearly identically for the 14~TeV LHC. We apply the same dimuon trigger, reconstruction criteria, and cuts. The higher luminosity may present challenges in the form of pile-up or higher reconstruction thresholds, but they are beyond the scope of our analysis.  The resulting sensitivity to $\Br(h\to aa \to b \bar b \mu^+ \mu^-)$ are shown in \figref{14tev}.

%%%%%%%%%%%%%%%%%%%%%%%%%%%%%%%%%
\subsection{Analysis with smaller jet radius}
\label{sec:SmallRadius}

A straightforward solution to the loss of sensitivity at low $m_a$ is to reduce the clustering radius of $b$-jets. Since $b$-tagging dominantly makes use of track-based information, and since the small-radius jets would not be used for triggering, there is in principle no obstacle to implementing such a modified $b$-tagger for a well-motivated analysis~\cite{Hobbs:2014pc}. The achievable $b$-tagging efficiencies should be comparable, and the use of smaller $b$-jets significantly improves sensitivity of $2b2\mu$ searches in the low $m_a$ regime.\footnote{For a recent theoretical discussion of small jet radius effects see~\cite{Dasgupta:2014yra}.}

We estimate the sensitivity possible with such a modified $b$-tagging system. The $Z^{(*)}/\gamma^*+2j$ background is regenerated with the same generator level cuts as for the conventional analysis in \S\ref{sec:ExpectedSensitivity}, except that we change the parton separation criterion from $\Delta R = 0.2$ to $0.1$. (No such requirement was imposed on the other generated backgrounds.) The resulting cross section for this background is shown in the last row of \tabref{background_xsec}. Jets are then clustered with a radius of $R = 0.2$ for both ATLAS and CMS, and the cut on  $\Delta R_{J_1J_2}$ is relaxed to be $> 0.2$.
Except for these two changes, we assume the analysis, including $b$-tagging efficiencies, proceeds identically as in \S\ref{sec:ExpectedSensitivity}. 

The dashed lines in ~\figref{8tev} and~\figref{14tev} show the resulting reach for 8 and 14 TeV. The sensitivity is significantly improved for $m_a \leq 20 \gev$. At higher masses, the $b$-jets are less collimated, and the smaller jet radius reduces the suppression of backgrounds compared to the conventional analysis, so that the sensitivity is reduced.  
A combination of both approaches therefore seems useful to achieve good sensitivity to all of the mass range.  
However, we will now show that a substructure analysis may have superior reach to the low $m_a$ region than 
the simple small-jet analysis presented in this subsection. 

\begin{figure*}[htbp]
   \centering
   \includegraphics[width=0.49\textwidth]{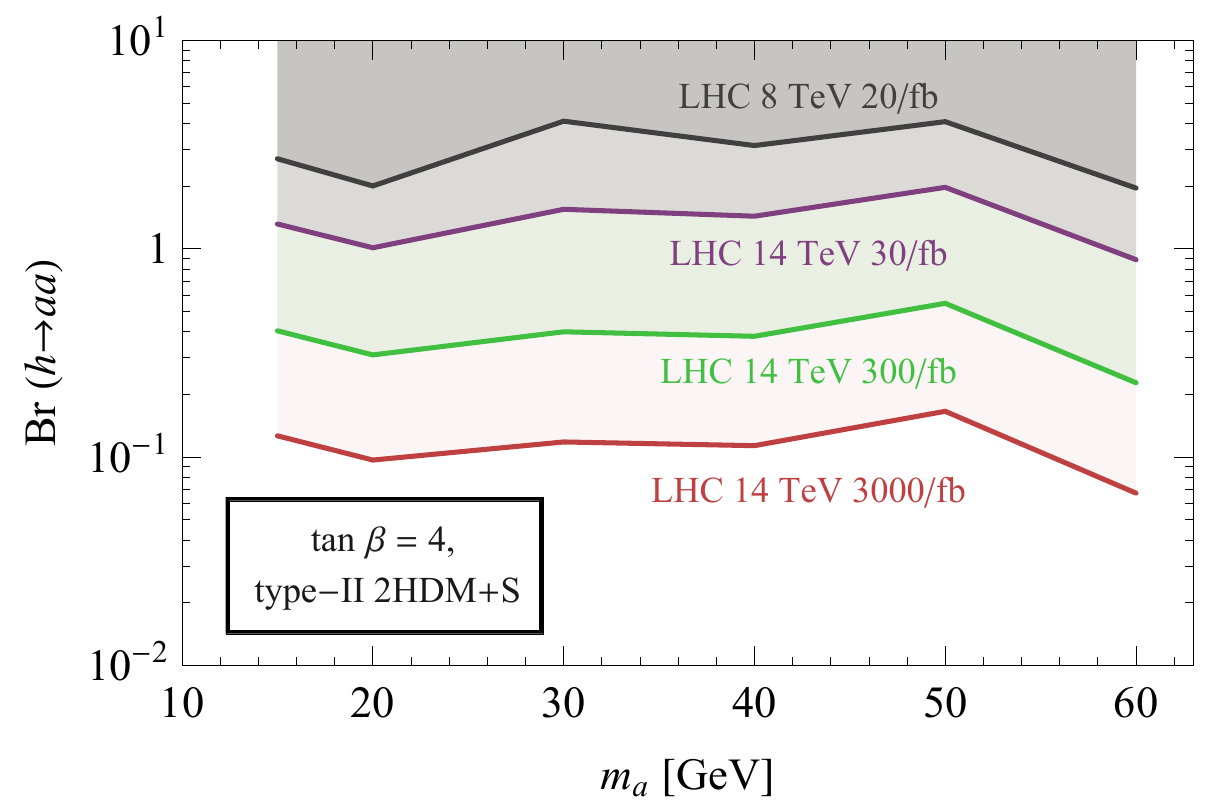}
   ~\includegraphics[width=0.49 \textwidth]{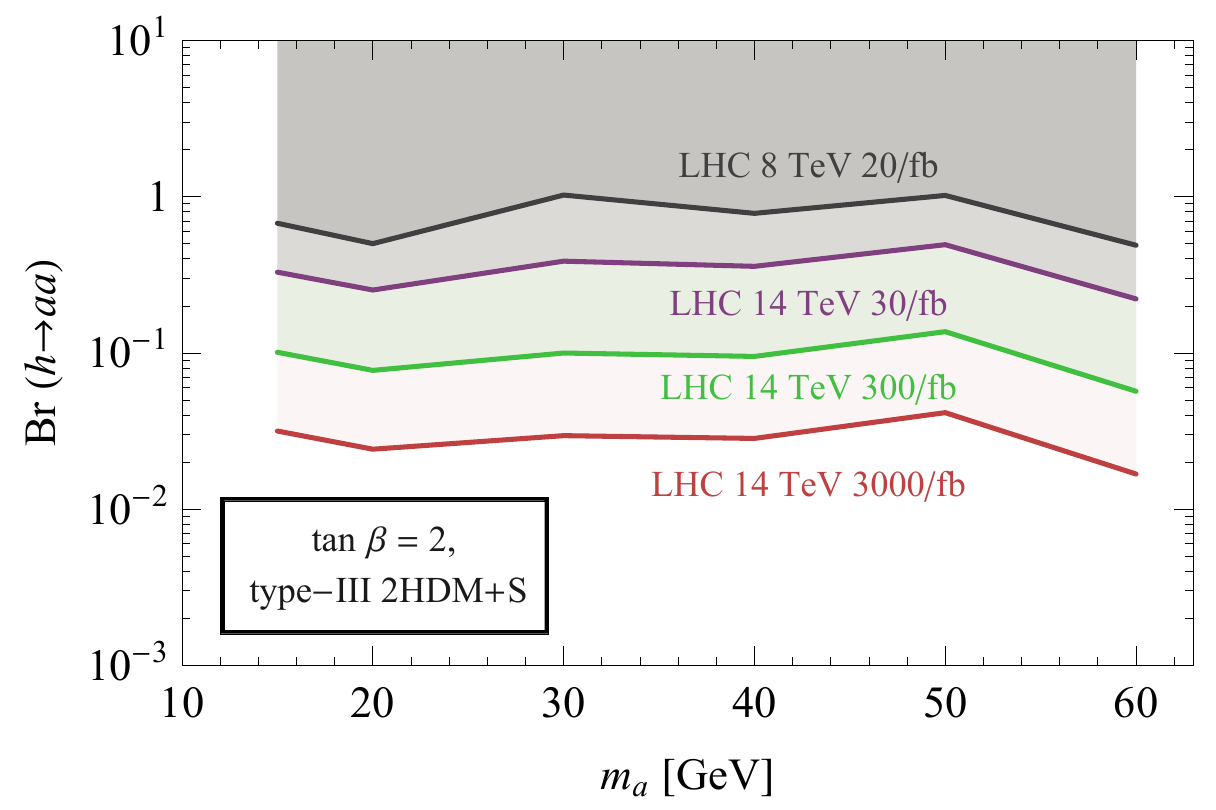} 
      \caption{
      Combined $95\%$ CL projected CMS sensitivities to $\Br(h\to aa)$ for the LHC at 8 and 14 TeV. To derive these sensitivities we need to make particular assumptions about how the scalar couples to the Standard Model fermions. 
      \emph{Left:} Type-II 2HDM+S with $\tan \beta = 4$, as in the left plot of \figref{Br-vs-tanbeta} in \S\ref{2HDMS}. The sensitivity to the SM+S model discussed in \S\ref{subsec:SMS} is almost identical.  \emph{Right:} Type-III 2HDM+S with $\tan \beta = 2$, as in the right plot of \figref{Br-vs-tanbeta} in \S\ref{2HDMS}. 
}
\label{fig:reaches}
\end{figure*}

%%%%%%%%%%%%%%%%%%%%%%%%%%%%%%%%%
\subsection{Jet substructure analysis}
\label{sec:JetSubstructures}

Sensitivity to the low-$m_a$ region can be further enhanced by making use of jet substructure techniques~\cite{Thaler:2010tr, Butterworth:2008iy, Plehn:2009rk, Krohn:2009th, Kribs:2009yh, Thaler:2008ju, Kaplan:2008ie}. The main goal is to increase signal acceptance without eroding background rejection.  Since the two $b$-jets from $a$ decays should be symmetric, we propose a jet substructure procedure based on the mass drop tagger (MDT)~\cite{Butterworth:2008sd}. 

The substructure analysis proceeds as follows. Triggered events satisfying the OS muons selection criterion are clustered into $R_f = 0.8$ fat jets with the Cambridge/Aachen (C/A) algorithm \cite{Dokshitzer:1997in, Wobisch:1998wt}. The (leading) fat jet is required to have one $b$-tag, and satisfy $p_T > 25 \gev$, $|\eta| < 2.5$.  We use the same $b$-tag efficiencies as in \S\ref{sec:ExpectedSensitivity}. We note that requiring {\it two} $b$-tags within the fat-jet will remove too much signal, as the $b$-tagged subjets need to presumably have a $p_T$ of at least 25~GeV to qualify as a proper subjet.  
If this threshold could be lowered, it would significantly improve sensitivity. 

We then analyze the substructure of the leading fat jet passing these criteria. The two hardest subjets, identified by undoing the last step of the C/A clustering, have to satisfy the MDT criteria
\begin{eqnarray}
 \mu&\equiv& \frac{\text{max}(m_{j_1}, m_{j_2})}{m_{j}} <0.67\,,
 \end{eqnarray}
 \begin{eqnarray}
 y&\equiv& \frac{\text{min}(p^2_{T\, j_1}, p^2_{T\, j_2})}{m_j^2} \Delta R^2_{j_1 j_2}> 0.09\,,
\end{eqnarray}
and $p_{T\,{j_{1,2}}} > 15 \gev$. We then apply the same $\Delta R$, missing energy, and invariant mass cuts for the two subjets and the two muons as in \S\ref{sec:ExpectedSensitivity}, with the exception of again relaxing the $\Delta R_{j_1j_2}$ cuts to $> 0.2$. 

The resulting 95\% CL sensitivities are shown as dot-dashed purple lines in \figref{8tev} and \figref{14tev} for 8~TeV and 14~TeV,  respectively.  The low-mass sensitivity is significantly enhanced compared to the previous two analysis approaches. Similarly to \secref{SmallRadius}, the conventional analysis does better at higher $m_a$ due to increased background rejection for an uncollimated signal. 

The impact of QCD multi-jet events with lepton fakes is hard to quantify for this substructure analysis without a data-driven study. For the resolved analyses, we found that muon fakes are reasonable to neglect if 0 or 2 $b$-tags are required. This substructure study requires only one fat-jet $b$-tag, but imposes strict kinematic requirements on its subjets. This may be enough to suppress multi-jet background, or it may be necessary to require both subjets to be $b$-tagged. As mentioned above, the $p_T$ threshold for $b$-tagging could weaken our projected sensitivity for small $m_a$, but determining whether this is necessary is beyond the scope of our analysis.

%%%%%%%%%%%%%%%%%%%%%%%%%%%%%%%%%%%%%%%%%%%%%%%%%%%%%%%%%%%%%%%%%%%%%
\section{Discussion}
\label{sec:discussion}

We have seen that combining the substructure and conventional analyses yields a fairly flat sensitivity of about 
$\mathrm{Br}(h \to 2a \to 2b2\mu) \lesssim 10^{-3}$
 for the 8 TeV LHC in the range $15 \gev \leq m_a \leq 60 \gev$. At 14 TeV with either 30, 300, and 3000 $\fb$ of data, the projected sensitivity increases to several times $10^{-4}$, $10^{-4}$, and several times $10^{-5}$, respectively. 
 
We can convert the projected reach on $\Br(h\to 2a \to 2b2\mu)$ to the projected reach on $\Br(h\to2a)$, but this is model-dependent.  
In a 2HDM+S model, for example, it depends on the Yukawa coupling type, see \S\ref{2HDMS}. In \figref{reaches}, we show the projected sensitivity to $\Br(h\to2a)$ from combining the substructure and conventional analyses for two 2HDM+S models, type-II with $\tan \beta = 4$ (very similar to SM+S) and type-III  with $\tan \beta = 2$.  In both cases, data at 14 TeV is required to meaningfully constrain exotic Higgs decays in these models, though in the latter case the 8~TeV constraint derived for $\mathrm{Br}(h\to2a)$ is less than 1. With the full HL-LHC (LHC at 14 TeV with $3000 \fb$) dataset, the exotic Higgs decay branching fraction can be constrained at the $10\%$ and $2\%$ level in these two scenarios, respectively. 

In motivating a $2b2\mu$ search, it is useful to compare its sensitivity to $\mathrm{Br}(h\to 2a)$ to that achievable in other channels. In particular:
\begin{itemize}
\item Earlier collider studies for the 14 TeV LHC~\cite{Kaplan:2011vf, Cao:2013gba} found $2\sigma$ sensitivity to $\mathrm{Br}(h\to 2a \to 4b) \approx 10\%$ with $300\fb$ of data. 
\item In constraining $\mathrm{Br}(h\to 2 a \to 4\tau)$ (and assuming Yukawa-ordered couplings, as we do here), the $(a\to 2\mu, a\to 2\tau)$ channel was found to be greatly superior to the $4\tau$ channel~\cite{Curtin:2013fra}. 
Depending on assumptions for reducible background, data from the LHC Run I can exclude $\mathrm{Br}(h\to2a\to4\tau) \lesssim 2 - 8\%$.
\item
A recent study of  $h\to 2a \to 2b2\tau$ decay from ggF  Higgs production~\cite{Bomark:2014gya}  claims considerably greater sensitivity to $\mathrm{Br}(h\to2a)$ in an NMSSM-like scenario than we find for $2b2\mu$. 
However, we find their study to be difficult to interpret, since it makes no attempt to incorporate trigger cuts.  In addition, highly optimistic $b$- and $\tau$-tag rates are assumed for
a low $p_T > 15 \gev$ threshold. 
The very tight mass windows employed also seem challenging at the LHC. 
For this reason, we will not consider their results in what follows, but the considered channel is interesting and deserves further study. \end{itemize}

Based on the existing theory-level studies done thus far, in a SM+S-like scenario (which generally includes the NMSSM and type-I 
and II 2HDM+S), the $4b$ search may be somewhat superior to $2b2\mu$, offering a sensitivity to $\mathrm{Br}(h\to2a)$ that is better by a factor of a few; the $4\tau/2\tau2\mu$ channel has no exclusion power. 
For more leptophilic scenarios, like the type-III 2HDM+S, the $2\tau2\mu$ search now constrains $\mathrm{Br}(h\to2a) \lesssim 10 - 40\%$ with LHC Run~I data~\cite{Curtin:2013fra}, performing much better than a $4b$ search. 
Here, the $2b2\mu$ channel should provide competitive sensitivity.  

The search for $h\to2b2\mu$ is therefore motivated for several reasons. Its sensitivity to the total exotic Higgs decay branching fraction is either competitive, or close to competitive, to searches involving $\tau$'s or only $b$'s. 
Apart from the complementarity of discovering new physics in several different channels, the much cleaner nature of the $2b2\mu$ signal makes our conclusions less reliant on the detailed modeling of $\tau$ and $b$-jet reconstruction at low $p_T$.  It could therefore turn out that $2b2\mu$ is the preferred channel in either leptophilic or NMSSM-type scenarios, although of course all the above-mentioned decay modes should be studied experimentally. Finally, although we did not consider this in detail here, it is also possible that $h\to X X'$ is the dominant exotic decay mode, where each scalar decays dominantly to $2b$ and $2\mu$, 
respectively (with e.g.~$X$ above the $2b$ threshold and $X'$ below the $2\tau$ threshold). 

%%%%%%%%%%%%%%%%%%%%%%%%%%%%%%%%%%%%%%%%%%%%%%%%%%%%%%%%%%%%%%%%%%%%%%
%%%%%%%%%%%%%%%%%%%%%%%%%%%%%%%%%%%%%%%%%%%%%%%%%%%%%%%%%%%%%%%%%%%%%
\section{Conclusion}
\label{sec:conclusion}

Exotic Higgs decays are uniquely sensitive to light scalars that are uncharged under the SM gauge groups.  We have demonstrated the sensitivity of a $h\to2a\to2b2\mu$ search for constraining theories with light scalars at the LHC.  This channel can arise naturally in models like the NMSSM or other 2HDM+S scenarios, as well as in general hidden valley scenarios. We performed a detailed collider analysis for an on-shell intermediate CP-odd scalar, though the results should be applicable to CP-even scalars as well, since we did not explicitly  use any angular information of the scalar decay. Already the 8 TeV LHC can probe $\mathrm{Br}(h\to2a) < 1$ in some 2HDM+S scenarios.  With its full dataset, the 14 TeV LHC will probe the exotic Higgs decay branching fraction to light scalars at the 1 - 10\% level. Depending on the details of soft $b$ and $\tau$ reconstruction, this sensitivity can be competitive or even superior to that offered by other channels that contain these final states.

For low intermediate scalar masses, a conventional resolved-jet analysis loses sensitivity due to the collimation of boosted $b$-jet pairs. Simply reducing the clustering radius of $b$-jets greatly enhances sensitivity in this region, but a dedicated jet substructure analysis may be even more powerful, fully exploiting the discovery potential for $m_a < 25 \gev$.

%%%%%%%%%%%%%%%%%%%%%%%%%%%%%%%%%%%%%%%%%%%%%%%%%%%%%%%%%%%%%%%%%%%%%%
%%%%%%%%%%%%%%%%%%%%%%%%%%%%%%%%%%%%%%%%%%%%%%%%%%%%%%%%%%%%%%%%%%%%%

\subsection*{Acknowledgements}
We thank Ze'ev Surujon for collaboration at the early stages of this project.  
We thank Matthew Strassler for many useful discussions about exotic Higgs decays, including in particular the $2b2\mu$ final state.  
We also thank Stefania Gori, John Hobbs,  Andrey Katz, Mariangela Lisanti, George Redlinger, Jessie Shelton, Scott Thomas, Dmitri Tsybychev  and Brock Tweedie for useful discussions. 
We are especially grateful to Stefan H\"oche and Steffen Schumann for consultations in setting up our \texttt{Sherpa} simulations. 
DC is supported in part by the NSF under Grants PHY-PHY-0969739 and PHY-1315155, and by the Maryland Center for 
Fundamental Physics. 
RE is supported by the DoE Early Career research program DESC0008061 and 
through a Sloan Foundation Research Fellowship.  
YZ is also supported through DoE grant DESC0008061. 

%%%%%%%%%%%%%%%%%%%%%%%%%%%%%%%%%%%%%%%%%%%%%%%%%%%%%%%%%%%%%%%%%%%%%
\appendix
\section{Estimation of the multi-jet QCD backgrounds}
\label{sec:MultiJet}

The high rate of QCD multi-jet processes means that the possibility of two QCD jets `faking' a pair of muons must be considered. This is a very rare process, occurring mostly due to heavy flavor decay inside of a jet with otherwise soft hadronic constituents that result in the muon passing isolation requirements. The rate for QCD jets resulting in a muon tag is estimated at $\sim 10^{-3}$ per $b/c$-jet~\cite{Sullivan:2010jk, Chatrchyan:2014aea}  and $\sim10^{-4}$ per light flavor jet~\cite{Curtin:2013zua}. 

These backgrounds are notoriously difficult to simulate in full detail. As pointed out  by~\cite{Sullivan:2006hb}, even large-scale full Monte Carlo simulations still lack the credibility to predict these fake muon backgrounds, and experimental analyses rely on data-driven methods to estimate their contributions. 

A framework for the simulation of fake leptons was proposed in  \cite{Curtin:2013zua}, in which differential mis-tag rates are derived from experimental information, then used to reweight event samples and hence obtain statistically reliable fake-lepton background distributions that preserve the kinematics of the source processes without simulating large numbers of events. This was successfully used to reproduce data-driven estimates of fake lepton backgrounds in \cite{Chatrchyan:2012sa, Chatrchyan:2012paa}. We will use this framework to very roughly estimate the size of QCD multi-jet background to our $2b2\mu$ search. Given the large uncertainties,  our estimate of the number of fake leptons should only be considered as qualitative.

We simulate QCD multi-jet backgrounds, together with the irreducible DY and $t \bar t$  backgrounds, at leading order and at parton level in  \texttt{MadGraph 5.1.14}~\cite{Alwall:2011uj}. We reweight the events using the procedure in~\cite{Curtin:2013zua}, then apply preselection cuts and compare the rates of multi-jet backgrounds to those of DY and $t \bar t$ backgrounds. Since the latter are included in our analyses (simulated to a much greater level of detail in \texttt{Sherpa 2.1.1}~\cite{Gleisberg:2008ta}), comparing irreducible to multi-jet backgrounds in this toy study will allow us to estimate the significance of lepton fakes to our analyses.

For the purpose of this estimate, we ignore the relatively small amount of momentum lost when the `jet' is reconstructed as a muon. We only need the mis-tag rate as a function of jet $p_T$. In~\cite{Curtin:2013zua}, this was parameterized by a simple linear function.
\beq
\epsilon_{j \to \mu} (p_{T_j}) =\epsilon_{200} \left[1-(1-r_{10})\frac{200-(p_{Tj}/\text{GeV})}{200-10}\right],
\eeq
where $\epsilon_{200}\!\equiv\! \epsilon_{j\to \mu} (200 \gev)$ and $r_{10}\!\equiv\! \epsilon_{j\to \mu} (10 \gev)/ \epsilon_{j\to \mu} (200 \gev)$. 
We adopt the three fake-rate benchmarks derived in~\cite{Curtin:2013zua} for the rate of a light jet faking a muon at the 8 TeV LHC:
\begin{enumerate}[(a)] \itemsep=0mm
\item $r_{10}=0$, $\epsilon_{200}=3.8\times 10^{-4}$\,; 
\item $r_{10}=0.5$, $\epsilon_{200}=1.6\times 10^{-4}$\,;
\item $r_{10}=1$, $\epsilon_{200}=0.85\times 10^{-4}$\,. 
\end{enumerate}
For the 14 TeV LHC, we adopt two benchmarks: 
\begin{enumerate}[(A)]\itemsep=0mm
\item the same parameters as (a);
\item $r_{10}=1$, $\epsilon_{200}=1.7\times 10^{-4}$. 
\end{enumerate}
For $b$/$c$-jets faking muons, we simply scale the mis-tag efficiency curve of the light jet by 
\bea
\epsilon_{b\to \mu} (p_{T\,b})&\approx& 50\times \epsilon_{j\to \mu}(p_{T\, j})\,,\\
\epsilon_{c\to \mu} (p_{T\,c})&\approx& 50\times \epsilon_{j\to \mu}(p_{T\, j})\,,
\eea
as suggested in~\cite{Sullivan:2010jk, Chatrchyan:2014aea}. (This may be pessimistic, see~\cite{Flowerdew:2007ef}.)

After reweighting the multi-jet events ($4j$, $4c$, $4b$, $2b2j$, $2b2c$, $2c2j$) according to these fake rate curves and applying preselection criteria, we find that irreducible DY backgrounds appear dominant when requiring zero or two $b$-tags. Therefore, for the analyses presented in Sec.~\ref{sec:ExpectedSensitivity} and \ref{sec:SmallRadius}, fake muon backgrounds can be safely ignored. For a single $b$-tag, fake background may be competitive with DY and $t \bar t$ after the preselection cuts, but adding that channel in any case does not improve sensitivity. 
For the jet-substrcture analysis presented in \S\ref{sec:JetSubstructures}, the fake background may be important, as there we require only one $b$-tag.  For this, a careful experimental study, using a data-driven background estimate, is required.  \

%%%%%%%%%%%%%%%%%%%%%%%%%%%%%%%%%%%%%%%%%%%%%%%%%%%%%%%%%%%%%%%%%%%%%

\bibliography{2b2mu.bib}
\bibliographystyle{h-physrev.bst}

\end{document}